\documentclass{JHEP3}

\newcommand{\GG}{{\cal{G}}}

\def\half{\frac{1}{2}}

\newcommand{\DD}{\mathcal{D}}   

\def\alp{\leavevmode\ifmmode {\alpha^\prime} \else ${\alpha^\prime}$ \fi}
\def\GN{G_{N}}

\newcommand{\ud}{\mathrm{d}}

\newcommand{\bea}{\begin{eqnarray}}
\newcommand{\beal}[1]{\begin{eqnarray}\label{#1}}
\newcommand{\eea}{\end{eqnarray}} 
\newcommand{\be}{\begin{equation}} 
\newcommand{\bel}[1]{\begin{equation}\label{#1}}
\newcommand{\ee}{\end{equation}} 
\newcommand{\rf}[1]{(\ref{#1})}
\newcommand{\nn}{\nonumber}
\newcommand{\bit}{\begin{itemize}}
\newcommand{\eit}{\end{itemize}}
\newcommand{\ben}{\begin{enumerate}}
\newcommand{\een}{\end{enumerate}}

\def\half{\frac{1}{2}}

\def\alp{\leavevmode\ifmmode {\alpha^\prime} \else ${\alpha^\prime}$ \fi}

\preprint{}

\title{On the gravity dual of strongly coupled charged plasma}

\author{Grzegorz Plewa\footnote{Email: g.plewa@ipj.gov.pl}\\
National Center for Nuclear Research, ul. Ho\.za 69, 00-681 Warsaw, Poland}

\author{Micha\l\ Spali\'nski\footnote{Email: mspal@fuw.edu.pl} \\
National Center for Nuclear Research, ul. Ho\.za 69, 00-681 Warsaw, Poland  \\
Physics Department, University of Bia{\l}ystok, ul. Lipowa 41, 15-424 Bia{\l}ystok, Poland}

\abstract{
  Locally asymptotically AdS solutions of Einstein equations coupled
  with a vector field with a weakly curved boundary metric 
  are found within the fluid-gravity gradient expansion up to second order in gradients. This geometry is 
  dual to $1+3$ dimensional hydrodynamics with a conserved current in a
  weakly curved background. The causal structure of the bulk geometry is
  determined and it is  shown that the black brane 
  singularity is shielded by an event horizon. 
}


\begin{document}

\section{Introduction}

Applications of the AdS/CFT correspondence to non-static geometries continue
to be an active area of research. This is motivated partly be the inherent
interest in non-equilibrium processes in strongly coupled quantum theories and
partly by pressing questions arising in this context in connection with
the phenomenology of heavy ion collisions \cite{CasalderreySolana:2011us}.

An important connection between slowly evolving geometries on the gravity side
and hydrodynamic states of N=4 supersymmetric Yang-Mills theory was uncovered
in \cite{Bhattacharyya:2008jc} following the earlier observations of
\cite{Janik:2005zt}. The simplest case of this connection involves only the
metric in the bulk, and describes the hydrodynamics of a fluid without any
conserved currents beyond the energy-momentum tensor. This so-called
fluid-gravity correspondence was subsequently generalized in various
directions
\cite{Banerjee:2008th,Erdmenger:2008rm,Bhattacharyya:2008ji,Bhattacharyya:2008mz}.

A very important case is that of hydrodynamics with conserved charges. Apart
from purely theoretical interest, the motivation for this arises in the
context of applications of relativistic hydrodynamics to the evolution of
quark-gluon plasma created in heavy ion collisions. In that case the conserved
charge is baryon number\footnote{In general one may also consider additional
  conserved charges, such as strangeness (as long as weak interactions are
  negligible).}. More generally, one is interested in the properties of gauge
theory plasma at finite temperature and chemical potential. Such a system is
dual to gravity interacting with a vector field. The generalization of
fluid-gravity duality to this situation was taken up in
\cite{Banerjee:2008th,Erdmenger:2008rm}, 
where a solution of the dual gravitational theory was described (up to second
order in the gradient expansion) and transport coefficients of the
charge-bearing plasma were calculated (see also
\cite{Hur:2008tq,Kalaydzhyan:2010iv}). Since the bulk gravity theory as it
appears in a string compactification\cite{Chamblin:1999tk} includes a
Chern-Simons term, the dual hydrodynamics includes terms unexpected from the
point of view of classical expositions of the subject. This has lead to very
interesting developments which clarify the effects of anomalies in
hydrodynamics \cite{Son:2009tf}.

The major goal of this article was to determine the causal structure of the
dual spacetime. This is motivated by two issues. First of all, the
gravitational solution found in the gradient expansion is smooth apart from a
singularity at $r=0$ (the black brane singularity). On general grounds it is
expected that this singularity should shielded by an event
horizon. Establishing this is important and nontrivial. The second issue is
that of hydrodynamic entropy currents
\cite{Loganayagam:2008is,Bhattacharyya:2008xc,Romatschke:2009kr}. The event
horizon defines an entropy current and the condition that its divergence be
non-negative imposes constraints on the transport coefficients\footnote{See
  however \cite{Jensen:2012jh,Banerjee:2012iz}}. It is natural to also consider
  entropy currents defined by dynamical horizons in the geometry
  \cite{Booth:2009ct,Booth:2010kr,Booth:2011qy}. This subject will be taken up
  elsewhere \cite{Plewa:2013rga}.

Much of this article is devoted to establishing the dual solution
itself. Compared to the earlier works the calculation presented here differs
in a number of ways. First of all, the results generalize the
findings of \cite{Banerjee:2008th,Erdmenger:2008rm} in that an arbitrary
weakly curved boundary metric is admitted. As in the case without charge, the
result can not be obtained just by covariantizing the flat-boundary result,
since there is an additional term which involves the curvature of the boundary
metric \cite{Bhattacharyya:2008ji}.  More importantly from the point of view
of locating the event horizon, the form of the solution obtained here is
somewhat simpler and more explicit than the earlier results. This is due to a
number of factors. Since one of the goals of this study was to determine the
causal structure of the dual spacetime geometry, it was convenient to use a
parameterization of the zeroth-order solution such that the event horizon at
that order can be expressed in a simple way. Such a parameterization (found in
\cite{Myers:2009ij}) also simplifies the gradient expansion, and especially
the perturbative determination of the event horizon of the slowly evolving
geometry. Another technical difference is that the present paper makes use of
a different gauge than that used in \cite{Banerjee:2008th,Erdmenger:2008rm},
following the choice made in
\cite{Bhattacharyya:2008ji,Bhattacharyya:2008mz}. The advantage of this gauge,
apart from simplicity, is that ingoing null geodesics are simply curves of
constant boundary coordinate $x$. This defines a natural bulk-boundary map,
which is an important element of the holographic construction of the
hydrodynamic entropy current.  Finally, explicit Weyl covariance in maintained
throughout (as in \cite{Bhattacharyya:2008ji,Bhattacharyya:2008mz}), which
simplifies the calculations as well as the form of the final results. It turns
out, that taken together these simplifications make it possible to write
completely explicit formulae for the metric and gauge field, which are somewhat
more complicated than those given in \cite{Bhattacharyya:2008mz} for the case
of uncharged plasma, but not significantly so. In particular, it is manifest
that the results given here reduce to those of \cite{Bhattacharyya:2008mz} in
the limit of vanishing charge.  Following the standard procedure of
holographic renormalization simple and complete expressions for the transport
coefficients are also obtained.
 
A very natural generalization of the computations presented here would be to
include a background gauge field (corresponding to background electric and
magnetic fields). This would lead to additional transport
coefficients. Calculations of this type have been done to first order in the
gradient expansion \cite{Hur:2008tq,Kharzeev:2011ds,Loganayagam:2011mu},
leading to results for the electrical conductivity, for example. In the
present study the extension to non-vanishing background fields was not made,
although carrying out such computations to second order is feasible. The
second order calculations described below are partly motivated by the fact that due to conformal
symmetry the differences between the event and apparent horizons only arise at
that order. While dynamical horizons are not considered in this article, the
methods used here could be used to find the apparent horizon for the geometry
under consideration, 

The structure of this article is as follows. Section \ref{thermo} reviews the
relevant static gravity solution and some aspects of its thermodynamics. 
Section \ref{genformsol} describes the general form of the solution and an
overview of the computation. Section \ref{ordbyord} presents the solution  up to
second order in the gradient expansion. Section \ref{holography} describes the
results of the 
holographic renormalization procedure and the results for the transport
coefficients. The causal structure of the geometry is studied in section
\ref{causal}, where the event horizons are located. Some closing remarks follow
in section \ref{closing}.

\section{The bulk theory, black branes and thermodynamics}
\label{thermo}

The action of the five-dimensional Einstein-Maxwell theory under consideration
reads
\beal{action}
S&=&\frac{1}{16\pi\GN}\int d^5x\sqrt{-g} \left(\frac{12}{L^2}
+ R - F^2 - \frac{4 \kappa}{3} \epsilon^{ABCDE} A_A F_{BC}F_{DE}\right)
\eea
where the cosmological constant is denoted by $12/L^2$. For this theory to be
a consistent truncation of type IIB
supergravity\cite{Chamblin:1999tk,Gauntlett:2006ai,Gauntlett:2007ma} the
Chern-Simons 
coupling $\kappa$ has to assume the value $1/2\sqrt{3}$. 

The action \rf{action} leads to the equations of motion
\bea
\label{eom}
\nonumber
G_{A B} - 6 g_{A B} + 2 F_{A C} F^{C}_{B} + \frac{1}{2} g_{A B} F_{C D}F^{C D}
&=& 0,
\\[0ex]
\nabla_{B} F^{A B} + \kappa \epsilon^{A B C D E} F_{B C} F_{D E} &=& 0.
\eea
Static black hole solutions possessing spherical symmetry as well as their
thermodynamics were discussed in detail in \cite{Chamblin:1999tk}. As shown there
(following \cite{Witten:1998zw}) a scaling limit gives rise to the following
solution:
\beal{bbsolution}
\nonumber
ds^2&=&- \frac{r^2 f(r)}{L^2}  dt^2+\frac{L^2}{r^2 f(r)}dr^2
+\frac{r^2}{L^2}(dx^2+dy^2+dz^2)\,,\\
A_t&=& h(r)\,,
\eea
where
\beal{func}
f(r) &=& \left(1-\frac{r_0^2}{r^2}\right)\left(1+\frac{r_0^2}{r^2}
-\frac{q^2}{r_0^2r^4}\right)\,,\\
h(r) &=&\frac{\sqrt{3} q}{2 L} \left(\frac{1}{r_0^2}-\frac{1}{r^2}\right)
\,.
\eea
This solution possesses a planar event horizon. In consequence of AdS/CFT
duality it describes the thermodynamics of plasma in equilibrium in flat
Minkowski space. 

The position of the (outer) event horizon is at $r=r_0$. There
is also an inner horizon at $r=r_-$, where 
\bel{rminus}
r_-^2=\half r_0^2\left(\sqrt{1+4\frac{q^2}{r_0^6}}-1\right)\,.
\ee
Using standard Euclidean techniques one finds the Hawking temperature of the
outer horizon \cite{Myers:2009ij} 
\bel{hawkingtemp}
T=\frac{r_0}{\pi L^2} \left(1-\frac{q^2}{2
r_0^6}\right)\,.
\ee
Note that the temperature vanishes for the extremal black hole with
$q^2/r_0^6=2$.

As discussed in \cite{Myers:2009ij} the chemical potential in the field theory
is related 
to the asymptotic behaviour of the gauge field. It can be expressed in terms
of $r_0$ and
the Hawking temperature by
 \be
r_0=\pi L^2 \frac{T}{2}\left(1+\sqrt{1+\frac{2}{3}\frac{\mu^2}{T^2}}
\right)\,.
 \ee

For the purposes of fluid-gravity duality it is appropriate to use coordinates
which are not singular at the event horizon. The Schwarzschild-like
coordinates used in \cite{Myers:2009ij} suffer from a coordinate singularity there. The
choice made in \cite{Bhattacharyya:2008jc} was 
to use Eddington-Finkelstein coordinates.
Starting from the coordinates used above one can use a
transformation of the form $r = r' + F(r')$  to reach such a gauge. In the
present case this 
results in the following expression of the charged black brane solution: 
\beal{bbsolutionef}
\nonumber
ds^2&=& 2 dr dt - \frac{r^2 f(r)}{L^2}  dt^2 + 
\frac{r^2}{L^2}(dx^2+dy^2+dz^2)\,,\\
A_t&=&   \frac{   \sqrt{3} q }{2 L r^2}  \, ,
\eea
where the function $f$ is given above in eq. \rf{func}.

\section{General form of the solution}
\label{genformsol}

\subsection{The gradient expansion}

The method of \cite{Bhattacharyya:2008jc} mimics the way relativistic
hydrodynamics arises from 
the static, thermodynamic description. The energy-momentum tensor of perfect
fluid hydrodynamics is just a boost of the equilibrium energy-momentum tensor,
where the temperature and boost parameters are allowed to depend on
position. Following the same idea one considers the
boost of \rf{bbsolutionef}, which describes equilibrium states, to some constant
velocity $u$, and then allows this velocity
and the temperature to depend on $x$. 
The boost parameter $u^{\mu}$ is a $4$-component velocity vector in the $x^{\mu}$
directions, normalized so that 
$u_{\mu} u^{\mu} = -1$ in the sense of the boundary metric
$h_{\mu \nu}$ (metric on the
conformal boundary of the locally asymptotically AdS spacetime \rf{bbb}).
Thus one is lead to consider the
geometry\footnote{From now on the constant $L$ is set to unity, and the
  notation is chosen to resemble that of reference
  \cite{Bhattacharyya:2008mz}.} 
\bel{bbb}
\textmd{d} s^2 = r^2 \left(P_{\mu \nu}-2 B  u_{\mu}u_{\nu} \right)
\textmd{d}x^{\mu} \textmd{d} x^{\nu} - 2 u_{\mu} \textmd{d}x^{\mu} \textmd{d} r,
\ee
where\footnote{The notation is chosen so that in uncharged limit ($q
  \rightarrow 0$, $\kappa \rightarrow 0$) $B$ is equal to $B(b r)$ as defined
  in \cite{Bhattacharyya:2008mz}.}  
\be
\label{bform}
B = \frac{1}{2} \left( 1-\frac{1}{b^4 r^4} \left( 1 + q^2 b^6 \right)+ \frac{q^2}{r^6}  \right)
\ee
and
\be
P_{\mu \nu} = h_{\mu \nu} + u_{\mu} u_{\nu}
\ee 
is the projector operator onto the space transverse to $u^{\mu}$, 
The vector potential takes the form:
\bel{ba}
A = \frac{ \sqrt{3} q }{2 r^2} u_{\mu} \textmd{d}x^{\mu}.
\ee
The constant parameter $b$ is just $1/r_0$ in the notation of the previous
section. The solution described there is recovered by going to the frame where
$(u^\mu)=(1,0,0,0)$. 

The geometry described above has a curvature singularity at $r = 0$. The
latter is shielded by the 
event horizon at $r = 1/b$. 
The parameter $b$ appearing in \rf{bbb} is related
to the Hawking temperature $T$ of the event horizon by eq. \rf{hawkingtemp},
which in the notation introduced above reads 
\be
T=\frac{1}{2 \pi b} (2 - q^2 b^6)\,.
\ee
The lines of constant $x^{\mu}$ in \rf{bbb}
are ingoing null geodesic, for large $r$ propagating in the direction set by
$u^{\mu}$, and the radial coordinate $r$ parameterizes them in an affine way
\cite{Bhattacharyya:2008xc}. 
Unlike black holes in asymptotically flat spacetime, the metric
\rf{bbb} supports perturbations varying much slower within
the transverse planes than within the radial direction. The parameter 
controlling the scale of variations in the radial
direction is $b$. 

The field configuration described above is a solution of the equations of
motion for constant $b, q, u^\mu$. If these parameters are allowed to depend
on $x$, the equations are violated by terms proportional to gradients of $b,
q, u$. To cancel these, so as to ensure that the fields still satisfy Einstein
equations, corrections need to be added to the metric and gauge potential
order by order in an expansion in the number of gradients.  Thus, if $b$,
$u^{\mu}$ and $h_{\mu \nu}$ are allowed to vary slowly compared to the scale
set by $b$, the metric \rf{bbb} should be an approximate solution of nonlinear
Einstein's equations with corrections organized in an expansion in the number
of gradients in the $x^{\mu}$ directions.  As in the uncharged case
\cite{Bhattacharyya:2008jc}, this turns out to be possible if and only if $b,
q, u$ satisfy differential equations which can be interpreted as the equations
of hydrodynamics \cite{Bhattacharyya:2008jc}.

Technically this can be done by considering an arbitrary point, say $x=0$, and
expanding the slowly varying quantities in Taylor series 
\beal{expan}
\nonumber
u^{\mu}(x) &=& {u^{\mu}}(0)+\epsilon \, x^{\alpha} \partial_{\alpha}
{u^{\mu}}(0)+ \dots \\
\nonumber
b(x) &=& b(0)+ \epsilon x^{\alpha} \partial_{\alpha} \, b(0) + \dots 
\nn\\
q(x) &=& q(0)+ \epsilon x^{\alpha} \partial_{\alpha} \, q(0) + \dots
\eea
Each 
derivative with respect to the ``boundary coordinates'' $x$ is tagged with a
power of $\epsilon$ for power counting purposes, and  $\epsilon$ is set to
unity at the end of calculations. 

The four-dimensional boundary metric components $h_{\mu \nu}$ are also
expanded around $x=0$, assuming that at zeroth order $h_{\mu \nu}(0) =
\eta_{\mu \nu}$. Moreover, to simplify computations it 
is very useful to adopt a locally geodesic coordinate system on the boundary,
so that all first order derivatives of $h_{\mu \nu}$ vanish. Thus one has 
\be
 h_{\mu \nu}(x) = \eta_{\mu \nu} + {\cal{O}}(\epsilon^2).
\ee
In general, the second order derivatives cannot of course be set to zero in
this way, but these contributions are guaranteed to be tensorial and so
the final results obtained are covariant in the boundary sense.

\subsection{Weyl covariance}

Weyl covariance in the bulk arises as an extension of the
conformal symmetry of N=4 supersymmetric Yang-Mills theory
\cite{Loganayagam:2008is,Bhattacharyya:2008mz}. A  
beautiful formalism allowing for manifest Weyl covariance in conformal
hydrodynamics was introduced by
Loganayagam \cite{Loganayagam:2008is} and applied to fluid-gravity duality in
\cite{Bhattacharyya:2008ji,Bhattacharyya:2008mz}.

The computation of the gradient expansion is simplified considerably by
adopting the approach of 
\cite{Bhattacharyya:2008mz}, which imposes at the outset the conditions of
Weyl invariance on 
the possible form of the solution. Conformal symmetry of the dual field theory
can be extended to the bulk as follows
\cite{Bhattacharyya:2008ji,Bhattacharyya:2008mz}: 
\bel{Weylrescalings}
g_{\mu \nu} \rightarrow e^{- 2\phi} g_{\mu \nu}, \quad u^{\mu} \rightarrow
e^{\phi} u^{\mu}, \ \quad b \rightarrow e^{-\phi} b \quad \mathrm{and} \quad r
\rightarrow e^{\phi} r 
\ee
where $\phi$ depends on the coordinates $x^{\mu}$
\cite{Bhattacharyya:2008mz}. A quantity which transforms homogeneously with a
factor of $e^{w\phi}$ is said to transform with Weyl weight $w$. The Weyl
weights of objects appearing in this paper are listed in  Appendix
\ref{appweyl}. 

The leading 
order metric \rf{bbb} is Weyl-invariant, but due to the presence
of $\ud r$ it  
does not retain its form at higher orders. It can however be written in a
manifestly Weyl-invariant form upon introducing a vector field
$\mathcal{A}_{\nu}$ defined by \cite{Loganayagam:2008is}
\bel{connectionFIELD}
\mathcal{A}_{\nu} \equiv u^\lambda\nabla_\lambda u_{\nu}-
\frac{\nabla_\lambda  u^\lambda}{3} u_{\nu} \, .
\ee
This quantity is of order one in the gradient expansion and transforms as a
connection under Weyl-transformations 
\bel{coonectionFIELDtrafo}
\mathcal{A}_{\nu} \rightarrow \mathcal{A}_{\nu} +\partial_{\nu}\phi \, .
\ee
A powerful tool for generating Weyl-covariant gradient terms is the
Weyl-covariant derivative $\mathcal{D}_{\mu}$, which uses the connection
$\mathcal{A}_{\mu}$ \rf{connectionFIELD} to compensate for derivatives of
the Weyl factor coming from derivatives of Weyl-covariant tensors. It has the
property that a Weyl-covariant derivative of a Weyl-covariant expression is
itself Weyl-covariant with the same weight
\cite{Loganayagam:2008is,Bhattacharyya:2008xc,Bhattacharyya:2008mz}.

The manifestly Weyl-invariant form of the metric \rf{bbb} reads
\bel{bbbw}
\textmd{d} s^2 = \left( r^2 P_{\mu \nu}-2r^2 B  u_{\mu}u_{\nu} \right)
\textmd{d}x^{\mu} \textmd{d} x^{\nu} - 2 u_{\mu} \textmd{d}x^{\mu}
\left(\textmd{d} r  + r\mathcal{A}_{\nu} \ud 
x^{\nu}\right).
\ee
The correction involving $ \mathcal{A}$ compensates an inhomogenous term in
the transformation of $dr$. It is of first order in gradients;
further contributions are needed for a complete solution at this order, but
they are by themselves Weyl-invariant. 

The static metric \rf{bbb} is a leading order approximation to a spacetime
whose metric is of the form\cite{Bhattacharyya:2008mz}
\bel{metricform}
\textmd{d} s^2 = \left( {\cal{G}}_{\mu \nu} - 2 u_{\mu} {\cal{V}}_{\nu} \right)
\textmd{d}x^{\mu} \textmd{d} x^{\nu} - 2 u_{\mu} \textmd{d}x^{\mu}
\left(\textmd{d} r  + r \mathcal{A}_{\nu} \ud  x^{\nu}\right),
\ee
with the condition $u^{\mu} \GG_{\mu \nu} = 0$  completely fixing the gauge
freedom \cite{Bhattacharyya:2008mz}. This implies a choice of gauge which is different than that of
\cite{Erdmenger:2008rm,Banerjee:2008th}. It is easy to see that lines of constant $x$ are geodesics
(affinely parameterized by $r$), as
in the case of the static metric. 

The metric \rf{metricform} is manifestly Weyl-invariant, provided that the
functions 
$\cal{V_{\mu}}$  are of unit Weyl weight and ${\cal{G}}_{\mu \nu}$ are Weyl
invariant. 
The simplest way to construct them is by summing individual tensorial
contributions of appropriate Weyl weight 
order by order in the gradient expansion, multiplied by 
scalar functions of the Weyl-invariant combinations $b r$ and $b^3 q$. 

Note finally that the Maxwell gauge field $A$ is a vector field
of Weyl weight zero.  It will be taken in the gauge $A_r=0$.

\section{The solution order by order}
\label{ordbyord}

\subsection{Zeroth order}

The solution at leading order is just the boosted charged black brane solution
\rf{bbb}. The metric is obviously of the form \rf{metricform} with
\bea
\nonumber
{\cal{V_{\mu}}} &= r^2 B  u_{\mu} 
\\
{\cal{G}}_{\mu \nu} &= r^2 P_{\mu \nu}
\eea
(strictly speaking, to retain only leading order terms, the term involving the
Weyl connection in \rf{metricform} must also be 
dropped, as it is of first order).

\subsection{First order}

To find the complete solution at first order one needs to classify the
possible terms that may appear. At first order there are:
\bit
\item no Weyl-invariant scalars
\item one Weyl-invariant pseudovector:
\be
l_{\mu} = \epsilon_{\mu\nu\lambda\rho} u^\nu \DD^\lambda u^\rho
\ee
\item one Weyl-invariant vector:
\be
{V_{0}}_\mu = q^{-1} P_{\mu}^{\nu} \DD_{\nu}{q}
\ee
\item one Weyl-invariant symmetric tensor\footnote{Symmetrization is defined
  as $A_{(\mu \, \nu)} = A_{\mu \nu}+ A_{\nu \mu}$.} of weight $w=-1$:
\be 
\sigma_{\mu \nu} = \half \DD_{(\mu} u_{\nu)}
\ee
\eit
Therefore the general form
of the Weyl-invariant structures appearing in the metric at this order is 
\bea
\nonumber
{\cal{V_{\mu}}} &=& r^2 B  u_{\mu} + r {F_{1}} l_{\mu} + b r^2 {F_0} {V_0}_{\mu},
\\
{\cal{G}}_{\mu \nu} &=& r^2 P_{\mu \nu} + 2 b r^2 {F_2} \sigma_{\mu \nu},
\eea
where $ {F_0}$, ${F_1}$ and ${F_2}$ are functions of the Weyl-invariants $r b$
and $b^3 q$. These functions are  
to be determined by solving the 
field equations up to linear order in gradients.
The factors of $b$ and $r$ above were chosen to ensure the
correct Weyl weights (and partly also for convenience - the choice made above
leads to a simple form of the differential equations). 

Similarly, Weyl covariance implies that the vector potential takes the form:
\bel{Afirst}
A = \left( \frac{ \sqrt{3} q u_{\mu}  }{2 r^2} + {Y_0} l_{\mu} + {\tilde{Y}_0}
{V_0}_{\mu} \right) 
\textmd{d}x^{\mu}.
\ee
Once again, ${Y_0}$ and ${\tilde{Y}_0}$  are functions of  $b r$ and $b^3 q$
and will be determined by solving the 
equations of motion. 

To find the unknown scalar functions $F_0,F_1,F_2,Y_0,\tilde{Y}_0$ the Ansatz described above
is inserted into the field equations and expanded in gradients using
\rf{expan}. This process, 
while tedious (thus best relegated to symbolic manipulation
software\footnote{For most calculations in this paper Mathematica was used,
  but for some the Cadabra package
  \cite{DBLP:journals/corr/abs-cs-0608005,Peeters:2007wn} was very convenient.}), 
is not significantly more complicated than what has to be done 
to reproduce the results of \cite{Bhattacharyya:2008mz}. 

As discussed in \cite{Bhattacharyya:2008jc}, the bulk field equations are of
two types: the 
constraint equations, which impose consistency conditions on the
slowly-varying background parameters $b(x), q(x), u^\mu(x)$ and the dynamical
equations which determine the functions which appear in the field Ansatz \rf{metricform}. 

The constraint equations take the form 
\bea
\label{constraints}
\nonumber
\partial_{0}q &=& - q \partial_{i}u_{i},\quad
(2-b^{6} q^2)\partial_{i}b-2 b (1+b^6 q^2) \partial_{0}{u_i} 
= q b^7 \partial_{i}q,  \\
\partial_{0} b &=& \frac{1}{3} b \partial_{i}u_{i}.
\eea
In terms of the Weyl-covariant derivatives
\bea
\nonumber
\DD_{\mu} b &=&  \partial_{\mu} b - {\cal{A}}_{\mu} b \label{Db}, \\
\DD_{\mu} q &=&  \partial_{\mu} q + 3 {\cal{A}}_{\mu} q \label{Dq}\ , 
\eea
one can rewrite the constraints \rf{constraints} as
\beal{focon}
\nonumber
(b^6 q^2 - 2) \DD_\mu b &=& q b^7 \DD_\mu q, \\
u^\mu \DD_\mu b &=& 0.
\eea
These equations can be interpreted as the
equations of hydrodynamics at order zero, which describe the perfect fluid
limit of the supersymmetric Yang-Mills plasma on the boundary. This will be discussed in more
detail in section \rf{holography} below. 

The remaining differential equations can be solved for the functions
$F_0,F_1,F_2,Y_0,\tilde{Y}_0$. Taking into consideration the requirements:
regularity across 
horizons and normalizability of the metric one can fix all constants of
integration to find
\bea
\nonumber
F_0 &=& -\frac{1}{2 b r}  +\frac{b^6
   q^2+2}{4 b^4 r^4}-\frac{q^2 \left(b^6 q^2+2\right)}{4 r^6 \left(b^6
  q^2+1\right)} \, +
\\[1ex]
\nonumber
  &+& \frac{(b^2 r^2 -1)(-b^6 q^2+b^2 r^2 + b^4 r^4)}{2 b^6 r^6}
 \int_{b r}^{\infty} \textmd{d}x \frac{x^4 (-b^6 q^2 (1+2 x)+x^2 (3+2 x+x^2))}{(1+x)^2 (-b^6 q^2+x^2+x^4)^2},
 \\[1ex]
\nonumber
{F_1} &=& \frac{\sqrt{3} b^4 \kappa  q^3}{r^5 \left(b^6 q^2+1\right)},
\\[1ex]
\nonumber
{F_2} &=& \int_{b r}^{\infty}  \textmd{d}x \frac{x(1+x+x^2)}{(1+x)(-b^6 q^2+x^2+x^4)} \, ,
\\[1ex]
\nonumber
{Y_0} &=& \frac{3 b^4 \kappa  q^2}{2 r^2 \left(b^6 q^2+1\right)},
\\[1ex]
\nonumber
\tilde{Y}_0 &=& - \frac{\sqrt{3} b q (2+ b^6 q^2)}{8 (1+b^6 q^2) r^2} +
\\
&+&
 \frac{\sqrt{3} b^3 q}{2} \int_{b r}^{\infty} \frac{\textmd{d} x}{x^3}
 \int_{x}^{\infty} \textmd{d}y \frac{y^4 (-b^6 q^2 (1+2 y)+y^2 (3+2
   y+y^2))}{(1+y)^2 (-b^6 q^2+y^2+y^4)^2} \, . 
\eea
The symbols $b$ and $q$ appearing above are understood as values at an
arbitrary point $x$, not necessarily $x=0$, which was just an irrelevant
choice made to implement the gradient expansion. As stressed in the original
papers on fluid-gravity duality \cite{Bhattacharyya:2008jc}, the equations are
ultralocal in $x$, which is 
the key feature which 
allows one to determine the solution. 

In the uncharged limit ($q \rightarrow 0 , \, \kappa \rightarrow 0$) one obtains
\be
F_2^{(q=0)} = \int_{b r}^{\infty} \frac{ x^3 - 1}{x (x^4-1)} \,  \textmd{d}x
\, ,
\ee
which is exactly the function $F(b r)$ appearing in the solution presented in
\cite{Bhattacharyya:2008mz}. In this limit the remaining functions vanish as
expected.

\subsection{Second order}

As in the previous section, the first step is to determine all the relevant
Weyl-invariant structures which may appear at second order. In doing this one
has to discard any terms which vanish or are 
not linearly independent after the first order constraints \rf{focon} are
taken into account. 

The results of this analysis are as follows:
\begin{itemize}
\item Scalars:
 \beal{scalars}
 \nonumber 
 S_1 &=& b^2 \sigma_{\mu \nu} \sigma^{\mu \nu},
 \\[0ex]
 \nonumber
 S_2 &=& b^2 \omega_{\mu \nu} \omega^{\mu \nu},
 \\[0ex]
 \nonumber
 S_3  &=& b^2 {\cal{R}},
 \\[0ex]
 \nonumber
 S_4 &=& b^2 q^{-2} P^{\mu \nu} \DD_{\mu}{q} \DD_{\nu}{q},
 \\[0ex]
 \nonumber
 S_5 &=& b^2 q^{-1} P^{\mu \nu} \DD_{\mu} \DD_{\nu}{q},
 \\[0ex]
 S_6 &=& b^2 q^{-1} P^{\mu \nu} l_{\mu} \DD_{\nu}{q} \, .
 \eea
 \item  Vectors:
 \bea
 \nonumber
 {V_1}_{\mu} &=& b P_{\mu \nu} \DD_{\rho} \sigma^{\nu \rho},
 \\[0ex]
 \nonumber
 {V_2}_{\mu} &=& b P_{\mu \nu} \DD_{\rho} \omega^{\nu \rho},
 \\[0ex]
 \nonumber
 {V_3}_{\mu} &=& b l^{\lambda} \sigma_{\mu \lambda},
 \\[0ex]
 \nonumber
 {V_4}_{\mu} &=& b q^{-1} \sigma_{\mu}^{\, \, \, \alpha} \DD_{\alpha}{q},
 \\[0ex]
   {V_5}_{\mu} &=& b q^{-1} \omega_{\mu}^{\, \, \, \alpha} \DD_{\alpha}{q} \, .
 \eea
 \item Tensors:
 \bea
 \nonumber
 {T_1}_{\mu \nu} &=& u^{\rho} \DD_{\rho} \sigma_{\mu \nu},
 \\[0ex]
 \nonumber
 {T_2}_{\mu \nu} &=& C_{\mu \alpha \nu \beta} u^{\alpha} u^{\beta},
 \\[0ex]
 \nonumber
 {T_3}_{\mu \nu} &=& \omega_{\mu}^{\, \, \, \lambda} \sigma_{\lambda \nu}+\omega_{\nu}^{\, \, \, \lambda} \sigma_{\lambda \mu},
 \\[0ex]
 \nonumber
 {T_4}_{\mu \nu} &=& \sigma_{\mu}^{\, \, \, \lambda} \sigma_{\lambda \nu} - \frac{1}{3} P_{\mu \nu} \sigma_{\alpha \beta} \sigma^{\mu \nu},
 \\[0ex]
 \nonumber
 {T_5}_{\mu \nu} &=& \omega_{\mu}^{\, \, \, \lambda} \omega_{\lambda \nu} + \frac{1}{3} P_{\mu \nu} \omega_{\alpha \beta} \omega^{\alpha \beta},
 \\[0ex]
 \nonumber
  {T_6}_{\mu \nu} &=&  \Pi_{\mu \nu}^{\alpha \beta} \DD_{\alpha} l_{\beta},
  \\[0ex]
  \nonumber
 {T_7}_{\mu \nu} &=& \frac{1}{2} \epsilon^{\alpha \beta}_{ \,\,\,\,\,\, \lambda (\mu } C_{\alpha \beta \nu) \sigma} u^{\lambda} u^{\sigma},
 \\[0ex]
 \nonumber
 {T_8}_{\mu \nu} &=& q^{-2} \Pi_{\mu \nu}^{\alpha \beta} \DD_{\alpha}{q} \DD_{\beta}{q},
 \\[0ex]
 \nonumber
 {T_9}_{\mu \nu} &=& q^{-1} \Pi_{\mu \nu}^{\alpha \beta} \DD_{\alpha} \DD_{\beta}q,
 \\[0ex]
 \nonumber
 {T_{10}}_{\mu \nu} &=& q^{-1} \Pi_{\mu \nu}^{\alpha \beta} l_{\alpha} \DD_{\beta}{q},
 \\[0ex]
  {T_{11}}_{\mu \nu} &=& \frac{1}{2} \epsilon_{(\mu}^{\, \, \, \, \, \alpha \beta \lambda}   \sigma_{\nu) \lambda}
  u_{\alpha} q^{-1} \DD_{\beta}{q} \, .
 \eea
\end{itemize}

Here $ \Pi_{\mu \nu}^{\alpha \beta} $ is the projector which can be used to
create symmetric, traceless tensors: 
\be
\Pi_{\mu \nu}^{\alpha \beta} = \frac{1}{2} \left( P_{\mu}^{\alpha}
 P_{\nu}^{\beta}+P_{\nu}^{\alpha} P_{\mu}^{\beta}-\frac{2}{3} P^{\alpha \beta}
 P_{\mu \nu} \right)\, 
\ee
The  scalar $\cal{R}$ is defined as in \cite{Bhattacharyya:2008mz} and
$C_{\mu \alpha \nu \beta}$ denotes the Weyl tensor:  
\be
C_{\mu \nu \lambda \sigma} = R_{\mu \nu \lambda \sigma} - \frac{1}{d-2} \left(
g_{\mu [\lambda} R_{\sigma] \nu}-g_{\nu [\lambda}R_{\sigma] \mu}  \right)+ 
\frac{1}{(d-1)(d-2)}  g_{\mu [\lambda}g_{\sigma] \nu} R \, .
\ee
On the basis of the above results one can write down the most general form for
the fields allowed by Weyl invariance:
  \bea
 \label{VGone}
 \nonumber
 {\cal{V}}_{\mu} &=& r^2 B u_{\mu} + r {F_1} l_{\mu}+b r^2 {F_0}{V_0}_{\mu}  + 
  r^2 \sum_{i=1}^{6} K_{i} S_{i}u_{\mu} + r \sum_{i=1}^{5} W_{i}
  {V_{i}}_{\mu}\, ,
 \\[0ex]
 {\cal{G}}_{\mu \nu} &=& r^2 P_{\mu \nu} + 2 b r^2 F_2 \sigma_{\mu \nu} + 
 r^2 \sum_{i=1}^{6} L_{i} S_i P_{\mu \nu} + \sum_{i=1}^{11}  H_{i} {T_i}_{\mu
   \nu}\, .
 \eea
The vector potential takes the form:
 \be
A = \left(   \frac{   \sqrt{3} q u_{\mu}   }{2 r^2} + {Y_0} l_{\mu}+ 
{\tilde{Y}_0} {V_0}_{\mu} +
 r \sum_{i=1}^{6} N_{i} S_i u_{\mu}   +  \sum_{i=1}^{5} Y_{i}{V_i}{_\mu}  
 \right) \textmd{d}x^{\mu} \, .
 \ee
The $39$ coefficient functions $K_i, L_i, N_i$ ($i=1,...,6$), $W_i, Y_i$
($i=1,...,5$), and ${H_i}$ ($i=1,...,11$) all depend on the Weyl invariant
variables $b^3 q$ and $b r$. 
Solving the equations of motion up to second
order determines these functions uniquely.
This procedure is more
cumbersome than at first order, but as before one can simplify it by first
establishing the constraints and the using them to simplify the remaining
equations. 
In this way all the functions appearing in \rf{metricform} are determined. The
integration constants, as at first order, are all fixed by conditions of
regularity, apart from the constants appearing in functions $K_{1},...,K{_6}$,
$Y_{1},...,Y_{5}$, which can be fixed by choice of 
frame. The results are listed in appendix \ref{appsecondorder}.

Moreover, from the constraint equations one obtains
relations involving second-order terms in $u_{\mu}$, $b$ and $q$. As in other
similar contexts, these coincide with the equations of hydrodynamics at order
one.

\section{Holography}
\label{holography}

The holographic dictionary of the AdS/CFT correspondence provides a
straightforward prescription for calculating physical quantities in the 
boundary theory. In practice, the computation involves subtraction of
divergences, which is done in a systematic way using holographic
renormalization \cite{Henningson:1998gx,Balasubramanian:1999re}. 

The expectation value of the energy momentum tensor is given by
\bel{holot}
T_{\mu \nu} = -\frac{1}{8 \pi \GN} \lim_{r \rightarrow \infty} r^2 \left(
K_{\mu \nu} - K H_{\mu \nu}+3 H_{\mu \nu} - E^{H}_{\mu \nu} \right), 
\ee
where $\GN$ is the $5$-dimensional Newton's constant, $H_{\mu \nu}$ is the
induced metric at the surface $r=const$ and $E^{H}_{\mu \nu}$ is the
corresponding   
Einstein tensor; $K_{\mu \nu}$ is the extrinsic
curvature and $K$ is its trace. 
Carrying out the calculation requires a careful evaluation of
the asymptotics of the integrals given in appendix \ref{appsecondorder}. One
then finds the following  
result  
\beal{emtensor}
 \nonumber
 T_{\mu \nu} &=& \frac{1}{16 \pi \GN} \Big( \, \frac{1+b^6 q^2}{b^4}(P_{\mu
   \nu}+3 u_{\mu} u_{\nu}) - \frac{2  \sigma_{\mu \nu}}{b^3}+ \frac{ 2 (1+c_1)
   {T_1}_{\mu \nu} }{b^2}+\frac{2 {T_2}_{\mu \nu}}{b^2}+\frac{2 c_1 \,
   {T_3}_{\mu \nu}}{b^2}+\nonumber\\
&+& \frac{2 {T_4}_{\mu \nu}}{b^2} 
+ \frac{4 b^4 q^2 (-1+b^6 q^2 (12
   \kappa^2-1))}{1+b^6 q^2} {T_5}_{\mu \nu}+ \frac{ 2 \sqrt{3} b^7 q^3 \kappa
 }{1+b^6 q^2} {T_6}_{\mu \nu} +\frac{c_8}{b^2} {T_8}_{\mu \nu} +\nonumber\\
&+&
 \frac{c_9}{b^2} {T_9}_{\mu \nu} + \frac{ c_{10}}{b^2} {T_{10}}_{\mu \nu} 
\, \Big),
 \eea
 where
 \bea
 \nonumber
 c_1 &=& -\frac{b^6 q^2+1}{2 } \int_{1}^{\infty} \frac{\textmd{d} x}{-b^6 q^2+x+x^2},
 \\[1ex]
 \nonumber
 c_8 &=&  \int_{1}^{\infty} \textmd{d} x \, {p_8}(x), \quad c_9 =
 \int_{1}^{\infty} \textmd{d} x \, {p_9}(x), \quad c_{10} =  \int_{1}^{\infty}
 \textmd{d} x \, {p_{10}}(x), 
 \eea
 with ${p_8}(x)$, ${p_9}(x)$, ${p_{10}}(x)$ are defined in the appendix
 \ref{appsecondorder} (together with the second order solutions).

The result \rf{emtensor} agrees with \cite{Bhattacharyya:2008mz} (for $d=4$) in the
limit $q \rightarrow 0$, $\kappa \rightarrow 0$. From it one can read off the 
values of the transport coefficients.

This formula \rf{emtensor} shows that the equations of state are $\epsilon=3 p$ (as
expected 
for a conformal theory) and 
\be
\epsilon =  \frac{3}{16 \pi \GN} \left(\frac{1}{b^4} + b^2 q^2  \right)
\ee
(which is also consistent with conformal symmetry). This formula shows an
interesting duality property, that is, invariance under the substitution 
\bea
\nonumber
b^2 &\longrightarrow&  \frac{1}{b q}, \\
q^2 &\longrightarrow& \frac{q}{b^3}.
\eea
Under this transformation the Weyl invariant quantity $b^3 q$ is inverted:
\be
b^3 q\longrightarrow \frac{1}{b^3 q}.
\ee
This duality can be easily expressed in terms of $T$ and $\mu$. 

Holographic renormalization of the current\footnote{The physical current, whose
  time component defines the density, differs from this by a factor of $2\pi$,
  due to the normalization of the chemical potential adopted earlier.}
proceeds according to the formula
\cite{Banerjee:2008th} 
\be
 J_\mu = \lim_{r \rightarrow \infty} \frac{r^2 A_\mu}{ 8 \pi \GN }.
 \ee
Inserting the solutions one finds:
  \bea
  J_\mu &=& \frac{1}{8 \pi \GN} \Big( \frac{\sqrt{3}q u_\mu}{2}
  + \frac{3 b^4 q^2 \kappa l_\mu}{2(1+b^6 q^2)}
  - \frac{\sqrt{3} b^3 q (2+ b^6 q^2)}{8 (1+b^6 q^2) b^2} {V_0}_\mu + \frac{3 \sqrt{3} b q }{8(1+b^6 q^2)}{V_1}_{\mu} +
 \nonumber\\[1ex]  
  &+&\frac{3 \sqrt{3}b^7 q^3 \kappa^2 }{(1+b^6 q^2)^2}{V_2}_{\mu} - \frac{3 b^4 \kappa  q^2}{2 \left(b^6 q^2+1\right)^2} {V_3}_{\mu}
  + \frac{2 {a_4} \left(b^6 q^2+1\right)+\sqrt{3} b^9 q^3}{16 b^2 \left(b^6
    q^2+1\right)^2} {V_4}_{\mu} +
 \nonumber\\[1ex]  
&+&
\frac{ {a_5} \left(b^6 q^2+1\right)+\sqrt{3} b^9 \left(24 \kappa ^2-1\right)
  q^3-\sqrt{3} b^3 q}{8 b^2 \left(b^6 q^2+1\right)^2} {V_5}_{\mu} 
  \Big).
 \eea
The constants $a_4$ and $a_5$ are 
 \bea
\nonumber
 a_4 &=& \int_{1}^{\infty} \textmd{d}x \frac{2 \sqrt{3} b^3 q \left(b^{12}
   q^4 \left(3-2 x^3\right)+b^6 q^2 x^2 \left(-3 x^2+2 x-3\right)+4
   x^3\right)}{x^2 \left(b^6 q^2-2\right) \left(b^6 q^2-x^2 
   \left(x^2+1\right)\right)},
   \\[1ex]
   \nonumber
   a_5 &=& -\frac{\sqrt{3} b^3 q \left(b^{12} q^4-3 b^6 q^2+2\right)}{2 b^6 q^2+2}+
   \int_{1}^{\infty} \textmd{d} x \, \frac{3 \sqrt{3} b^3 q {F_0} \left(b^6
     q^2 \left(5 x^2-9\right)+x^2 \left(3 x^4+5\right)\right)}{x^4}. 
\eea
This formula contains the values of the charge transport coefficients.

 \section{Causal structure of the dual geometry}
\label{causal}

This section is devoted to locating the event horizon in the geometry
discussed above. It is important to verify that the metric 
singularity at $r=0$ is shielded by an event horizon, as expected on general
grounds. As explained in \cite{Bhattacharyya:2008xc} (see also
\cite{Booth:2010kr,Booth:2011qy}) is possible to 
find the event horizon in the gradient expansion by taking advantage of the
fact that at zeroth order its location is known. 

To determine the event horizon it is simplest to assume that it can be presented
as the level set of a scalar function $S(r,x)$. This function must be
Weyl-invariant and can be written in the gradient expansion in terms of all
the independent scalars available up to some order. Since the event horizon at
order zero is at $r=1/b$, one has 
\bel{formofs}
S(r,x) = b(x) r - g(x)
\ee
where $g(x)$ is a Weyl-invariant function expanded in gradients of $b, q$: 
\be
g(x) = g_0(x) + g_1(x) + g_2(x) \ + \dots 
\ee
Here $g_k$ denotes a linear combination of all Weyl-invariant scalars at
order $k$ in the gradient expansion. Thus $g_0(x)$ is a function of Weyl
weight zero, i.e. it depends only on 
the combination $b^3 q$. There are no Weyl-invariant scalars a order 1, and 6 
at order 2, so one expects to find 
\bea
\nonumber
g_0(x) &=& \lambda(b^3 q)\\
\nonumber
g_1(x) &=& 0 \\
g_2(x) &=& \sum_{k=1}^{6} h_k(b^3 q) S_k \, ,
\eea
where the $S_i$ are the $6$ independent Weyl-invariant scalars \rf{scalars}. 
The functions $h_i$ and $\lambda$ will be 
determined in due course. Once this is
done, the expression for the 
position of the event horizon will take the form
\bel{horpos}
r_H = \frac{1}{b} \left(\lambda +  \sum_{k=1}^{6} h_k S_k 
\right) \, . 
\ee
The normal covector
to a surface of the form \rf{formofs} is 
\bel{ds}
m = dS \, , 
\ee
which up to second order in the gradient expansion is
\bel{masform}
m = r\ db +  b\ dr + d\lambda\, .
\ee
It is convenient to write the normal in terms of the Weyl-covariant 
derivatives (given in \rf{Db}, \rf{Dq}). 
One then has\footnote{Here $\lambda'\equiv\lambda'(b^3 q)$.}
\be
m =  \left(r \DD_\mu b + \lambda' \left(b^3 \DD_\mu q + 3 q b^2 \DD_\mu
b\right) \right) dx^\mu 
+  b \left( dr + r \mathcal{A}_\mu dx^\mu\right) \, . 
\ee
Since the event horizon is a null surface this normal must satisfy $m^2 =
0$. At leading order this condition has two solutions: $\lambda$ = 1, which is
the outer event horizon at order zero, and $\lambda=r_-$, where $r_-$ is given
in \rf{rminus}, which is the inner horizon of the charged black brane. 
At order two the condition fixes the functions $h_k$ appearing in
\rf{horpos}. Carrying out this calculation with the metric obtained earlier
one finds the outer event horizon at
\bea
\label{outer hor position}
r &=& \frac{1}{b}+\frac{ 3 \left(b^6 q^2-2\right) K_1 \left(b^3
  q,1\right)+1}{3 b \left(b^6 q^2-2\right)^2} S_1+ 
 \frac{2 b^{18} \left(12 \kappa ^2-5\right) q^6+b^{12} \left(25-48 \kappa
   ^2\right) q^4+80 b^6 q^2+45}{60 b \left(b^6 q^2-2\right) \left(b^6
   q^2+1\right)^2} S_2 +\nn \\ 
 &-& \frac{ 1 }{  12 b ( 2- b^6 q^2  )  } S_3  
 + \Big( \frac{{K_4}\left(b^3 q,1\right)}{b \left(b^6
   q^2-2\right)}+\frac{b^{11} q^4 \left(17 b^{12} q^4+28 b^6 q^2+20\right)}{32
   \left(b^6 q^2-2\right)^3 \left(b^6 q^2+1\right)^2} \Big) S_4 + \nn\\
& +& 
  \Big( \frac{{K_5}\left(b^3 q,1\right)}{b \left(b^6 q^2-2\right)}-\frac{b^5
    q^2 \left(b^6 q^2+2\right)}{4 \left(b^{18} q^6-3 b^{12} q^4+4\right)}
  \Big) S_5 \,  + \nn
 \\
 &+& \frac{4 \left(b^{18} q^6-3 b^6 q^2-2\right) {K_6}\left(b^3
   q,1\right)+\sqrt{3} b^9 \kappa  q^3 \left(b^6 q^2+4\right) \left(3 b^6
   q^2+2\right)}{4 b \left( - b^{12} q^4 +b^6 q^2+2\right)^2} S_6.
\eea
It can be checked directly, by taking the $q\rightarrow 0$ limit, that the above
formula reduces to the correct expression obtained in the uncharged case in
\cite{Bhattacharyya:2008xc} (see also \cite{Bhattacharyya:2008mz,Booth:2011qy}).

Formally, the inner horizon at second order is given by
\bel{innerhor}
r= r_{-} + h_1 S_1 + h_2 S_2 +h_3 S_3+h_4 S_4 + h_5 S_5 + h_6 S_6,
\ee
where the coefficient functions $h_i$ can be found in appendix
\ref{appinnerhor}. As seen from the expressions given there, these functions
(given in the form of integrals) diverge at $r=r_-$. This is not surprising,
since the corrections to the metric and gauge field are divergent there
already at first order, indicating a breakdown of the gradient expansion. 

It may also seem interesting to consider the extremal limit, in which $r_{-} =
1/b $ so the inner and outer event horizons at 
zeroth order coincide. This corresponds to $b^3 q = \sqrt{2} $. The gradient 
corrections to this horizon include divergent integrals, so one must conclude
that the gradient construction breaks down in this extremal limit
\cite{Erdmenger:2008rm}.

\section{Conclusions}
\label{closing}

The main results of this article are 
the determination of the geometry dual
to hydrodynamics with a conserved current in an arbitrary weakly curved
background and the determination of its causal structure. The calculations of
the bulk field theory solution in the gradient expansion 
described here follow the pattern of earlier calculations of this type and have
benefited from a number of technical insights accumulated in this field. This
made it possible to present the dual description in a
form suitable for studying its causal structure following the method described in
\cite{Bhattacharyya:2008xc} (see also \cite{Booth:2010kr}). The location of
the event horizon is found explicitly in section \rf{causal}, thus showing
that the locus $r=0$ is not a naked singularity. As is well known, the
event horizon can also be used to define a hydrodynamic entropy current for the
dual field theory \cite{Bhattacharyya:2008xc}.

Apart from the event horizon, the charged black brane geometry possesses also
an inner horizon. The gradient corrections to the metric (as well as the gauge field)
are singular at the location of the zeroth order inner horizon, indicating a
breakdown of the gradient expansion. 
It is straightforward to formally determine the location of the inner horizon in the
gradient expansion. The double integrals appearing in
the formal expression for its location are divergent. This is reminiscent of the remarks in
\cite{Erdmenger:2008rm} concerning the extremal limit, in which the inner and outer horizons
coincide. As noted in \cite{Erdmenger:2008rm}, the solution obtained in the
gradient expansion develops  
singularities as the temperature $T$ vanishes with $q$ nonzero. This is
apparently an indication that this limit is not described by hydrodynamics on
the field theory side of the duality
\cite{Gubser:2000ec,Bhattacharyya:2007vs}. It could however be that the
extremal limit could be taken in such a way that an alternative gradient
expansion would be valid. 

An obvious extension of this work would be to determine the entropy current
associated with the event horizon. This is left to a future work
\cite{Plewa:2013rga}, 
where the entropy currents associated with dynamical horizons will also be
discussed.

Finally, it would also be interesting to include background fields in this
calculation, along the lines of \cite{Hur:2008tq,Kharzeev:2011ds,Loganayagam:2011mu},
where some important partial results are obtained. 

\begin{acknowledgments}
The authors would like to thank Micha\l\ P. Heller for discussions and helpful
comments on the manuscript, and Kasper
Peeters for his excellent package {\tt Cadabra}
\cite{DBLP:journals/corr/abs-cs-0608005,Peeters:2007wn}. This work was
partially supported by Polish Ministry of Science and Higher Education grant
\emph{N N202 173539}.

\end{acknowledgments}


\appendix

\section{Second order functions}
\label{appsecondorder}

In the expressions below the prime denotes a partial derivative with respect to $br$. 

\subsection{Scalar sector}
 
\bea
  \nonumber
  N_1  & =& q \int_{b r}^{\infty} \textmd{d}x \,  \Big( \frac{b^2 x}{2 \sqrt{3} r}-\frac{\sqrt{3} x^3}{2 r^3}+\frac{x^4}{\sqrt{3} b r^4} \Big)
 \frac{ (1+x+x^2)^2 }{ (1+x)^2 (-b^6 q^2 + x^2 + x^4 )^2 }
   ,
   \\[1ex]
   \nonumber
   N_2 &=& -\frac{3 \sqrt{3} b^6 \kappa ^2 q^5}{5 r^9 \left(b^6 q^2+1\right)^2}-\frac{\sqrt{3} q}{4 b^2 r^5}-\frac{\sqrt{3} b^2 \kappa ^2 q^3}{r^7
   \left(b^6 q^2+1\right)},
   \\[1ex]
   \nonumber
   N_3  &=& 0,
   \\[1ex]
   \nonumber
   N_4 &=& \int_{b r}^{\infty} \textmd{d}x \Big( \frac{\sqrt{3} b^5 {F_0} q^3 x^2 \left(b^6 q^2 (2 x-3)+4 x-3\right) (x-b r)^2 (b r+2 x)}{r^4 (x-1)^2 (x+1)^2 \left(b^6 q^2+1\right) \left(-q^2 b^6
   +x^4+x^2\right)^2} \, +
   \\[1ex]
   \nonumber
   & +& \frac{2 \sqrt{3} {F_0}^2 q x^3 \left(b^6 q^2 \left(x^2-2\right)+x^2\right) (x-b r)^2 (b r+2 x)}{b r^4 (x-1)^2 (x+1)^2 \left(-q^2 b^6 +x^4+x^2\right)^2} \, +
   \\[1ex]
   \nonumber
   & -&\frac{b^5 q^3 x (x-b r)^2 (b r+2 x)}{8 \sqrt{3} r^4 (x-1)^2 (x+1)^3 \left(b^6 q^2+1\right)^2 \left(-q^2 b^6 +x^4+x^2\right)^3} \Big( 
   \\[1ex]
   \nonumber
 &+& 2 (x-1)^2 x^2 \left(2 x^4+18 x^3+21 x^2+18 x+7\right)+b^{18} q^6 \left(-4 x^4+2 x^3+9 x^2+x-14\right)+
 \\[1ex]
 \nonumber
 & +& b^{12} q^4 \left(4 x^8+10 x^7-17 x^6-3 x^5-15 x^4+15 x^3+35 x^2+17 x-28\right)+
 \\[1ex]
 \nonumber
 &+& b^6 q^2 \left(8 x^8+32 x^7-49 x^6-21 x^5-33 x^4+15 x^3+34 x^2+16 x-14\right)
    \Big)
          \Big),
  \\[1ex]
  \nonumber
 N_5 &=& \int_{b r}^{\infty} \textmd{d} x \frac{q x \left(b^6 q^2 (2 x+1)-x^2 \left(x^2+2 x+3\right)\right) (x-b r)^2 (b r+2 x)}{4 \sqrt{3} b r^4 (x+1)^2 \left(-q^2 b^6 +x^4+x^2\right)^2},
 \\[1ex]
 \nonumber
  N_6 &=& \int_{b r}^{\infty} \textmd{d} x \Big( -\frac{24 b^8 {F_0} \kappa  q^4 (x-b r)^2 (b r+2 x)}{r^4 (x-1) x^3 (x+1) \left(b^6 q^2+1\right) \left(-q^2 b^6 +x^4+x^2\right)} +
\\[1ex]
\nonumber
&-& \frac{b^2 \kappa  q^2 (x-b r)^2 (b r+2 x)}{2 r^4 (x-1) x^4 (x+1)^3 \left(b^6 q^2+1\right)^2 \left(-q^2 b^6 +x^4+x^2\right)^3} \Big( b^{24} q^8 \left(x^5+2 x^4+3 x^2+24 x+18\right)+
\\[1ex]
\nonumber
&+& b^6 q^2 x^4 \left(-9 x^9-2 x^8-19 x^7-9 x^6+11 x^5+7 x^4+9 x^3+3 x^2+4 x+5\right)
\\[1ex]
\nonumber
&-& 2 x^7 \left(3 x^6+2 x^5+x^4-3 x^2-2 x-1\right)
\\[1ex]
\nonumber
&+& b^{18} q^6 \left(2 x^9-4 x^8-7 x^7-18 x^6-48 x^5-45 x^4-54 x^3-48 x^2+12 x+18\right)+
\\[1ex]
\nonumber
&+& b^{12} q^4 x^2 \Big(-3 x^{11}+2 x^{10}-5 x^9+15 x^8+43 x^7+47 x^6+60 x^5+57 x^4+15 x^3+
\\[1ex]
\nonumber
&+& 6 x^2-18 x-27\Big)
 \Big)  
                  \Big),\\
   \nonumber
 L_1  &=& \frac{2}{3} {F_2}^2 - \frac{2}{3} \int_{b r}^{\infty} \frac{
   \textmd{d} x}{x^2} \int_{x}^{\infty} \textmd{d}y \, y^2 ({F_2}')^2, 
   \\[1ex]   
   \nonumber
L_2  &=& \frac{1}{3 b^2 r^2}-\frac{4 b^6 \kappa ^2 q^4}{5 r^6 \left(b^6 q^2+1\right)^2},
   \\[1ex]
   \nonumber
  L_3  &=& 0,
   \\[1ex]
   \nonumber
  L_4 &=&  \int_{b r}^{\infty} \textmd{d}x \Big(-\frac{4 b^3  q x^4 \left(b^6 q^2+1\right)}{3 (x-1) (x+1) \left(b^6 q^2-2\right) \left(b^6 q^2-x^2 \left(x^2+1\right)\right)} \frac{\partial F_0}{\partial (b^3 q)} + 
\\[1ex]
\nonumber
  &+&  \frac{4 {F_0}^2 x^5 \left(b^6 q^2 \left(2 x^2-3\right)+2 x^2\right)}{3 (x^2-1)^2 \left(-b^6 q^2  +x^4+x^2\right)^2}+
 \frac{{F_0} x^4}{3 (x^2-1)^2 (-b^6 q^2+x^2+x^4)} \Big( 2 x^2 (x^4-1)+ 
\\[1ex]
\nonumber
&+&   b^{18} q^6 (7 x^2+6 x-15)+b^{12} q^4 (x^6+12 x^2-5)+b^6 q^2 (3 x^6+3 x^2-24 x+10) \Big)+
   \\[1ex]
   \nonumber
   &+& \frac{b^6 q^2 x^3   }{12 (x^2-1)^2 \left(b^6 q^2-2\right) \left(b^6 q^2+1\right)^2 \left(b^6 q^2-x^2 \left(x^2+1\right)\right)^2}( b^{18} q^6 \left(2 x^3-3 x^2+16 x-16\right)+
   \\[1ex]
   \nonumber
   &-& 2 b^{12} q^4 \left(x^7+3 x^2-22 x+16\right)+2 b^6 q^2 \left(x^7-3 x^3+6 x^2+2 x-8\right)+4 x \left(x^6-x^2+6 x-6\right)  )
   \\[1ex]
   \nonumber
   & -&\frac{2 x^4 {N_4}''}{3 \sqrt{3} b^3 q}-\frac{10 x^3 {N_4}'}{3 \sqrt{3} b^3 q}-\frac{2 x^2 {N_4}}{\sqrt{3} b^3 q}
       \Big),
  \\[1ex]
   \nonumber
   L_5  &=& -\int_{b r}^{\infty} \textmd{d}x \Big( \frac{2 x^4 {F_0}}{3 \left(x^2-1\right) \left(-q^2 b^6 +x^4+x^2\right)}+\frac{x^3 \left(b^6 q^2 (x-2)+2 (x-1)\right)}{6
   \left(x^2-1\right) \left(b^6 q^2+1\right) \left(b^6 q^2-x^2 \left(x^2+1\right)\right)}
   +
   \\[1ex]
   \nonumber
   &+&\frac{2 x^4 {N_5}''}{3 \sqrt{3} b^3 q}+\frac{10 x^3 {N_5}'}{3 \sqrt{3} b^3 q}+\frac{2 x^2 {N_5}}{\sqrt{3} b^3 q}
    \Big),
   \\[1ex]
   \nonumber
   L_6  &=& \int_{b r}^{\infty} \textmd{d}x \Big( \frac{8 \sqrt{3} b^9 {F_0} \kappa  q^3}{(x-1) x (x+1) \left(b^6 q^2+1\right) \left(b^6 q^2-x^2 \left(x^2+1\right)\right)}
   -\frac{2 x^4 {N_6}''}{3 \sqrt{3} b^3 q}-\frac{10 x^3 {N_6}'}{3 \sqrt{3} b^3 q}-\frac{2 x^2 {N_6}}{\sqrt{3} b^3 q}+
   \\[1ex]
   \nonumber
   & -& \frac{b^3 \kappa  q}{\sqrt{3} (x-1) x^2 (x+1)^2 \left(b^6 q^2+1\right)^2 \left(b^6 q^2-x^2 \left(x^2+1\right)\right)^2} \Big(  
   \\[1ex]
   \nonumber
   &+& 4 x^5 \left(x^6+x^5+x^4-x^2-x-1\right)+b^{18} q^6 \left(2 x^4+2 x^3+4 x^2-7 x-13\right)  +
\\[1ex]
\nonumber
&-& b^{12} q^4 \left(2 x^8+6 x^7+12 x^6-7 x^5-15 x^4-15 x^3-21 x^2+x+13\right)+
   \\[1ex]
   \nonumber
   & +& b^6 q^2 x^2 \left(4 x^9+4 x^8+4 x^7-2 x^6-10 x^5-22 x^4-3 x^3+7 x^2+7 x+11\right)
   \Big) \Big)
   \\[1ex]
  \nonumber
  K_1  &=& 
- \frac{1}{2 b^2 r^2} + \frac{1}{b^4 r^4} \int_{b r}^{\infty} \textmd{d} x \Big( -x + \frac{x(1+x+2 x^2 -b^6 q^2 (1+x))}{6(1+x)(-b^6 q^2+x^2+x^4)} +
\\[1ex]
\nonumber
& -& \frac{2 b^6 q^2}{x^3} \int_{x}^{\infty} \textmd{d}y \,  y ({F_2}')^2  - \frac{x^2-3 x^6 +b^6 q^2 (x^2-6)}{3 x^4} \int_{x}^{\infty} \textmd{d}y \,  y^2 ({F_2}')^2  \Big) ,
 \\[1ex]
 \nonumber
K_2  &=&
\frac{1}{2 b^2
   r^2}-\frac{6 b^6 \kappa ^2 q^6}{5 r^{12} \left(b^6 q^2+1\right)^2}-\frac{5 q^2}{12 b^2 r^8}-\frac{7 b^2 \kappa ^2 q^4}{5 r^{10} \left(b^6 q^2+1\right)}+
\\[1ex]
\nonumber
&+& \frac{5 b^{18} q^6+3 b^{12} \left(4
   \kappa ^2+5\right) q^4+15 b^6 q^2+5}{20 b^6 r^6 \left(b^6 q^2+1\right)^2},
   \\[1ex]
   \nonumber
   K_3 &=& 
 \frac{1}{  12 b^2 r^2  },
 \\[1ex]
 \nonumber
 K_4  &=& -\frac{b^5 q^2}{3 r \left(4 b^6 q^2+4\right) }+ \frac{1}{b^4 r^4} \int_{b r}^{\infty} \textmd{d}x \Big( \frac{-b^6 q^2 x}{48 \left(x^2-1\right)^2 \left(b^6 q^2+1\right)^2 \left(b^6 q^2-x^2 \left(x^2+1\right)\right)^2} \Big( 
 \\[1ex]
 \nonumber
 &+& b^{18} q^6 \left(8 x^5+3 x^4-20 x^3-2 x^2+12\right)+
\\[1ex]
\nonumber
&-& 4 x^2 \left(2 x^7+15 x^6-18 x^5+x^4-2 x^3-3 x^2+6 x-1\right)+
 \\[1ex]
 \nonumber
&+& b^{12} q^4 \left(-8 x^9-15 x^8+36 x^7-4 x^6+24 x^5+15 x^4-64 x^3-12 x^2+24\right)+
\\[1ex]
\nonumber
&-& 4 b^6 q^2 \left(4 x^9+15 x^8-27 x^7+2 x^6-6 x^5-6 x^4+17 x^3+3 x^2-3\right)   \Big)+
\\[1ex]
\nonumber
&+& \frac{x^4 {F_0}}{6 \left(x^2-1\right)^2 \left(b^6 q^2+1\right) \left(-q^2 b^6 +x^4+x^2\right)^2} \Big( b^{18} q^6 \left(3 x^2+3 x-7\right)+
\\[1ex]
\nonumber 
 &+& b^{12} q^4 \left(-12 x^6-9 x^5+21 x^4+9 x^2+9 x-14\right)+3 \left(3 x^{10}-4 x^6+x^2\right)+
 \\[1ex]
 \nonumber
&+& b^6 q^2 \left(9 x^{10}-24 x^6-18 x^5+21 x^4+9 x^2+6 x-7\right)
 \Big)+
 \\[1ex]
 \nonumber
 &+& \frac{ {F_0}^2 x^5 \left(b^{12} q^4 \left(2 x^2-3\right)+b^6 q^2 \left(-12 x^6+15 x^4+4 x^2-3\right)+2 \left(3 x^{10}-6 x^6+x^2\right)\right)}{3 \left(x^2-1\right)^2
   \left(-q^2 b^6 +x^4+x^2\right)^2}+
   \\[1ex]
   \nonumber
  & -& \frac{x^4 \left(b^6 q^2-3 x^4+1\right) {N_4}''}{3 \sqrt{3} b^3 q}-\frac{x \left(b^6 q^2 \left(5 x^2+6\right)-15 x^6+5 x^2\right) {N_4}'}{3 \sqrt{3} b^3
   q}
\\[1ex]
\nonumber
&-& \frac{{N_4} \left(b^6 q^2 \left(x^2+2\right)-3 x^6+x^2\right)}{\sqrt{3} b^3 q} \Big),
   \\[1ex]
   \nonumber
   K_5  &=& \frac{b^6 q^2+2}{3 b r \left(4 b^6 q^2+4\right)}+\frac{1}{b^4 r^4} 
   \int_{b r}^{\infty} \textmd{d} x \, \Big( \frac{x \left(x^2 \left(3 x^3+3 x^2+3 x+1\right)-b^6 q^2 \left(x^2+2 x+2\right)\right)}{6 (x+1) \left(-q^2 b^6 +x^4+x^2\right)} + 
   \\[1ex]
   \nonumber
   &-& \frac{x^4 \left(b^6 q^2-3 x^4+1\right) {N_5}''}{3 \sqrt{3} b^3 q}-\frac{x \left(b^6 q^2 \left(5 x^2+6\right)-15 x^6+5 x^2\right) {N_5}'}{3 \sqrt{3} b^3
   q}+
\\[1ex]
\nonumber
&-& \frac{ {N_5} \left(b^6 q^2 \left(x^2+2\right)-3 x^6+x^2\right)}{\sqrt{3} b^3 q} \Big),
   \\[1ex]
   \nonumber
  K_6  &=& -\frac{2 b^2 \kappa  q}{\sqrt{3} r \left(b^6 q^2+1\right)} + \frac{1}{b^4 r^4} \int_{b r}^{\infty} \textmd{d} x \, \Big( \frac{-b^3 \kappa  q}{2 \sqrt{3} (x-1) x^2 (x+1)^2 \left(b^6 q^2+1\right)^2 \left(b^6 q^2-x^2 \left(x^2+1\right)\right)^2} \cdot
  \\[1ex]
  \nonumber
  & \cdot& \Big( b^{24} q^8 \left(2 x^4+2 x^3+x^2-4 x-7\right)+4 x^5 \left(3 x^7+x^6+x^5+x^4-3 x^3-x^2-x-1\right)+
\\[1ex]
\nonumber
&+& b^{18} q^6 \left(-8 x^8-6 x^6+10 x^5+35 x^4+14 x^3+10 x^2-5 x-14\right)+
\\[1ex]
\nonumber
&+& b^6 q^2 x^2 \left(18 x^{10}+2 x^9+32 x^8+5 x^7-35 x^6-11 x^5-23 x^4-7 x^3+7 x^2+7 x+5\right) + 
\\[1ex]
\nonumber
&+& b^{12} q^4 \Big(6 x^{12}-2 x^{11}+13 x^{10}-14 x^9-46 x^8-22 x^7+
\\[1ex]
\nonumber
&-& 40 x^6-8 x^5+40 x^4+19 x^3+14 x^2-x-7\Big)
   \Big)
   \\[1ex]
\nonumber
&+& \frac{2 \sqrt{3} b^9 \kappa  q^3 {F_0} \left(b^6 q^2-5 x^4+1\right)}{x \left(x^2-1\right) \left(b^6 q^2+1\right) \left(b^6 q^2-x^2
   \left(x^2+1\right)\right)}-\frac{{N_6} \left(b^6 q^2 \left(x^2+2\right)-3 x^6+x^2\right)}{\sqrt{3} b^3 q}+
   \\[1ex]
   \nonumber
   &-&\frac{x^4 \left(b^6 q^2-3 x^4+1\right) {N_6}''}{3 \sqrt{3} b^3 q}-\frac{x \left(b^6 q^2 \left(5 x^2+6\right)-15 x^6+5 x^2\right) {N_6}'}{3 \sqrt{3} b^3 q} \Big)
  \eea

 \subsection{Vector sector}
 
\bea
\nonumber
   Y_1 &=& 
 -\sqrt{3} b^3 q \int_{b r}^{\infty}  \frac{\textmd{d}x}{x^3} \int_{x}^{\infty} \textmd{d}y \frac{y^6 (2+y) }
 {(1+y)^2 (-b^6 q^2 +y^2 + y^4)^2}+\frac{3 \sqrt{3} b q}{8  \left(b^6 q^2+1\right)r^2},
  \\[1ex]
  \nonumber
  Y_2  &=& \frac{3 \sqrt{3} b^7 \kappa ^2 q^3}{ \left(b^6 q^2+1\right)^2 r^2},
  \\[1ex]
  \nonumber
  Y_3  &=& \frac{3 b^6 \kappa  q^2}{b^6 q^2+1} \int_{b r}^{\infty}  \frac{\textmd{d}x}{x^3} \int_{x}^{\infty} \textmd{d}y
  \frac{ y^3 \left(b^6 q^2 (y+2)+3 y^5+6 y^4+9 y^3+6 y^2+4 y+2\right)}{(y+1)^2  \left( -q^2 b^6 +y^4+y^2\right)^2} +
\\[1ex]
\nonumber
&-& \frac{3 b^4 \kappa  q^2}{2 r^2 \left(b^6 q^2+1\right)^2},
\\[1ex]
\nonumber
Y_4  &=& - \int_{b r}^{\infty}  \frac{\textmd{d}x}{x^3} \int_{x}^{\infty} \textmd{d}y
\frac{y^7}{(1-y^2)^2 (-b^6 q^2 +y^2+y^4)} 
\int_{1}^{y} \textmd{d}z \frac{2 \sqrt{3} b^3 q (z-1)}{z^2 \left(b^6 q^2-2\right) \left(b^6 q^2-z^2 \left(z^2+1\right)\right)} \Big(
\\[1ex]
\nonumber
 && 3 b^{12} q^4 \left(z^2+z+1\right)-b^6 q^2 z^2 \left(z^4+z^3+2 z^2-z+3\right)+2 z^3 \left(z^3+z^2+2 z+2\right) \Big)+
\\[1ex]
\nonumber
&+& \frac{2 {a_4} \left(b^6 q^2+1\right)+\sqrt{3} b^9 q^3}{16 b^2 r^2 \left(b^6 q^2+1\right)^2},
 \\[1ex]
 \nonumber
 Y_5  &=& \frac{\sqrt{3} b^3 q}{b^6 q^2+1} \int_{b r}^{\infty}  \frac{\textmd{d}x}{x^3} \int_{x}^{\infty} \textmd{d}y
\frac{y^7}{(1-y^2)^2 (-b^6 q^2 +y^2+y^4)} 
\int_{1}^{y} \frac{  \textmd{d}z }{z^5} \Big( 
\\[1ex]
\nonumber
&& b^{12} q^4 \left(-2 z^2+9 z-10\right)+2 z^2 \left(z^4-1\right)+2 b^6 q^2 \left(z^6-2 z^2+9 z-5\right)+
\\[1ex]
\nonumber
&+& 3 z {F_0} \left(b^{12} q^4 \left(5 z^2-9\right)+b^6 q^2 \left(3 z^6+10 z^2-9\right)+z^2 \left(3 z^4+5\right)\right) \Big)+
\\[1ex]
\nonumber
&+& \frac{\sqrt{3} {a_5} \left(b^6 q^2+1\right)+3 b^9 \left(24 \kappa ^2-1\right) q^3-3 b^3 q}{8 \sqrt{3} b^2 r^2 \left(b^6 q^2+1\right)^2}
\eea
\bea
   \nonumber
  W_1  &=&  - b r \int_{b r}^{\infty} \textmd{d} x \Big( \frac{(x-1) (x+2)}{(x+1) \left( -q^2 b^6+x^2 +x^4\right)}+
\\[1ex]
\nonumber
&-& \frac{2 \left(b^6 q^2 \left(2 x^2-3\right)+2 x^2\right)}{x^7} \int_{x}^{\infty}\textmd{d}y \frac{y^6 (2+y)}{(1+y)^2 (-b^6 q^2+y^2+y^4)^2} \Big)+
  \\[1ex]
  \nonumber
  &-&\frac{3 \left(b^6 q^2 r^2-b^4 q^2+r^2\right)}{4 b^3 r^5 \left(b^6 q^2+1\right)},
   \\[1ex]
   \nonumber
   W_2 &=& 
-\frac{1}{2 b r}+\frac{6 b^7 \kappa ^2 q^4}{r^5 \left(b^6 q^2+1\right)^2},
   \\[1ex]
   \nonumber
   W_3  &=& b r \int_{b r}^{\infty} \textmd{d}x \Big( 
-\frac{\left(b^6 q^2 r^2-b^4 \left(q^2+r^6\right)+r^2\right) {Y_3}''}{\sqrt{3} b^4 q r^3}+\frac{\left(b^6 q^2 r^2-3 b^4
   \left(q^2-r^6\right)+r^2\right) {Y_3}'}{\sqrt{3} b^5 q r^4}+
   \\[1ex]
   \nonumber
   &-&\frac{\sqrt{3} \kappa  q \left(b \left(b r^2 \left(3 b^6 q^2+2\right)+r \left(3 b^6 q^2+2\right)+2 b^5 q^2-3 b^5 r^6-3 b^4 r^5-3 b^3 r^4-b^2 r^3\right)+2\right)}{b^2 r^3
   \left(b^6 q^2+1\right) (b r+1) \left(b^4 q^2-b^2 r^4-r^2\right)}+
\\[1ex]
\nonumber
&+& \frac{8 \sqrt{3} {F_2} \kappa  q}{b^2 r^5}
   \Big),
\\[1ex]
   \nonumber
   W_4 &=&  b r \int_{b r}^{\infty} \textmd{d}x \Big( 
   \frac{\left(b^6 q^2 r^2-3 b^4 \left(q^2-r^6\right)+r^2\right) {Y_4}'}{\sqrt{3} b^5 q r^4}-\frac{(b r-1) (b r+1) \left(b^4 q^2-b^2
   r^4-r^2\right) {Y_4}''}{\sqrt{3} b^4 q r^3} +
   \\[1ex]
   \nonumber
   & \quad -&\frac{2 b^2 r^4 {F_0} }{b^6 q^2 r^2-b^4 \left(q^2+r^6\right)+r^2}-\frac{2 r^3 \left(b^8 q^2+b^2\right)-r \left(b^6 q^2+2\right)}{2 b \left(b^6
   q^2+1\right) (b r-1) (b r+1) \left(b^4 q^2-b^2 r^4-r^2\right)}
   \Big),
   \\[1ex]
   \nonumber
   W_5 &=&  b r \int_{b r}^{\infty} \textmd{d}x \Big( 
   -\frac{(b^2 r^2-1) \left(b^4 q^2-b^2 r^4-r^2\right) {Y_5}''}{\sqrt{3} b^4 q r^3}+\frac{\left(b^6 q^2 r^2-3 b^4
   \left(q^2-r^6\right)+r^2\right) {Y_5}'}{\sqrt{3} b^5 q r^4}+
\\[1ex]
\nonumber   
   & \quad +&\frac{1}{2 b^3 r^5 \left(b^6 q^2+1\right) \left(b^2 r^2-1\right) \left(b^4 q^2-b^2 r^4-r^2\right)} \Big( 48 b^4 \kappa ^2 q^2 \left(b^6 q^2 r^2-b^4 \left(q^2+r^6\right)+r^2\right)+
   \\[1ex]
   \nonumber
   & \quad +& r^2 \left(-2 \left(b^6 q^2 r+r\right)^2-3 b^5 q^2 r \left(b^6 q^2+2\right)+2 b^4 r^6 \left(b^6 q^2+1\right)+6 b^4 q^2 \left(b^6 q^2+1\right)\right) 
   \Big)+
   \\[1ex]
   \nonumber
   & \quad -&\frac{ {F_0} \left(5 b^6 q^2 r^2-b^4 \left(7 q^2+r^6\right)+5 r^2\right)}{b^8 q^2 r^4-b^6 r^2 \left(q^2+r^6\right)+b^2 r^4}
   \Big)
\eea

 \subsection{Tensor sector}
 
\bea
 \nonumber
 H_1 &=& 
 - (b r)^2 \int_{b r}^{\infty} \textmd{d} x \frac{x }{-b^6 q^2 + x^2+x^4} \left(1+\frac{1}{1-x^2}
 \int_{1}^{x} \textmd{d}y \left( 6 y^2 F_2+4 y^3  {F_2}'   \right)  \right),
   \\[1ex]
   \nonumber
   H_2 &=& 
- 2 (b r)^2 \int_{b r}^{\infty} \textmd{d}x \frac{x }{-b^6 q^2+x^2+x^4},
   \\[1ex]
   \nonumber
   H_3 &=& 
(b r)^2 \int_{b r}^{\infty} \textmd{d} x \frac{x}{-b^6 q^2 + x^2+x^4} \left(1-\frac{1}{1-x^2}
 \int_{1}^{x} \textmd{d}y \left( 6 y^2 F_2+4 y^3  {F_2}'   \right)  \right),
 \\[1ex]
 \nonumber
 H_4 &=&
2 (b r)^2 {F_2}^2- 2 (b r)^2 \int_{b r}^{\infty} \textmd{d}x \frac{x }{-b^6 q^2+x^2+x^4},
   \\[1ex]
   \nonumber
   H_5 &=& 
\frac{2(b r)^2}{(1 + b^6 q^2)^2} \int_{b r}^{\infty} \textmd{d} x \frac{1}{x^7 (-b^6 q^2 + x^2 + x^4)} \Big(
 -(1+b^6 q^2)^2 x^4 \left( x^2+x^4+b^6 q^2 (1+2 x^2) \right)+
 \\[1ex]
 \nonumber
& +& 12 b^{12} \kappa ^2 q^4 \left(b^6 q^2 \left(2 \left(x^6+x^4+x^2\right)+3\right)+ x^2 (x^2-1) \left(2 x^2+1\right)\right)   \Big),
   \\[1ex]
   \nonumber
   H_6  &=& 
 \frac{2 \sqrt{3} b^{11} \kappa  q^3 r^2 }{1+b^6 q^2}  \int_{b r}^{\infty} \textmd{d}x \frac{x^2+x+1}{x^2 (x+1)  \left(-b^6 q^2+x^2+x^4\right)},
   \\[1ex]
   \nonumber
   H_7  &=& 0,
   \\[1ex]
  \nonumber
  H_8 &=& -(b r)^2 \int_{b r}^{\infty} \textmd{d}x \frac{x }{(1-x^2)(-b^6
    q^2+x^2+x^4)} \int_{1}^{x} \textmd{d}y \,  p_{8}(y), 
     \\[1ex]
  \nonumber
  H_9 &=& -(b r)^2 \int_{b r}^{\infty} \textmd{d}x \frac{x }{(1-x^2)(-b^6
    q^2+x^2+x^4)} \int_{1}^{x} \textmd{d}y \,  p_{9}(y), 
     \\[1ex]
  \nonumber
  H_{10} &=& -(b r)^2 \int_{b r}^{\infty} \textmd{d}x \frac{x }{(1-x^2)(-b^6
    q^2+x^2+x^4)} \int_{1}^{x} \textmd{d}y \,  p_{10}(y), 
       \\[1ex]
  \nonumber
  H_{11} &=& 0,
 \eea
 where
 \bea
\nonumber
 {p_8}(x) &=&   -\frac{4 b^3  q x^2 \left(b^6 q^2+1\right) \left(b^6 q^2 \left(x^2-3\right)+3 x^6+x^2\right)}{\left(x^2-1\right) \left(b^6 q^2-2\right) \left(b^6 q^2-x^2 \left(x^2+1\right)\right)}  \frac{\partial F_0}{\partial(b^3 q)} +
 \\[1ex]
 \nonumber
 &+& \frac{4 {F_0}^2 x^3 \left(b^{12} q^4 \left(4 x^4-15 x^2+12\right)+b^6 q^2 x^2 \left(3 x^4+8 x^2-15\right)+4 x^4\right)}{\left(x^2-1\right)^2 \left( -q^2 b^6 +x^4+x^2\right)^2}+
 \\[1ex]
 \nonumber
 &+& \frac{2 {F_0} x^2 }{\left(x^2-1\right)^2 \left(b^6 q^2-2\right) \left(b^6 q^2+1\right) \left( -q^2 b^6 +x^4+x^2\right)^2} \Big(  4 x^8+2 x^4 +
 \\[1ex]
 \nonumber
 & +& b^{24} q^8 \left(2 x^4+9 x^3-22 x^2-12 x+24\right)+4 b^6 q^2 x^2 \left(4 x^6+3 x^5-6 x^4+2 x^2-9 x+4\right)
\\[1ex]
\nonumber
&-& 6 x^{12}+b^{18} q^6 \left(8 x^8-3 x^7-6 x^6+8 x^4+9 x^3-28 x^2+6\right)+
 \\[1ex]
 \nonumber
 &+& 2 b^{12} q^4 \left(3 x^{12}+10 x^8-15 x^6+6 x^4-18 x^3+5 x^2+24 x-9\right)
 \Big)
 \\[1ex]
 \nonumber
 &-& \frac{b^6 q^2 x}{4 \left(x^2-1\right)^2 \left(b^6 q^2-2\right) \left(b^6 q^2+1\right)^2 \left(b^6 q^2-x^2 \left(x^2+1\right)\right)^2 } \Big( 
 \\[1ex]
 \nonumber
&& b^{24} q^8 \left(3 x^4-12 x^3+8 x^2+36 x-36\right)+8 x^2 \left(3 x^6-6 x^5+x^4-3 x^2+6 x-1\right)+
\\[1ex]
\nonumber
 &+& b^{18} q^6 \left(-3 x^8-36 x^7+44 x^6+9 x^4-72 x^3+28 x^2+132 x-96\right)+
 \\[1ex]
 \nonumber
 &+& 4 b^6 q^2 \left(3 x^8-24 x^7+15 x^6-9 x^4+12 x^3+23 x^2-12 x-6\right)+
\\[1ex]
\nonumber 
 &-& 6 b^{12} q^4 \left(x^8+14 x^7-16 x^6+x^4+10 x^3-14 x^2-8 x+14\right)
        \Big),
 \\[1ex] 
 \nonumber
  {p_9}(x) &=& \frac{x}{\left(x^2-1\right) \left(b^6
   q^2+1\right) \left(b^6 q^2-x^2 \left(x^2+1\right)\right)} \Big( b^6 q^2 \left(b^6 q^2 (3 x-4)+6 x-4\right)+
\\[1ex]
\nonumber
&+& 2 x {F_0} \left(b^{12} q^4 \left(x^2-3\right)+b^6 q^2 \left(3 x^6+2 x^2-3\right)+3 x^6+x^2 \right) \Big),
 \\[1ex]
 \nonumber
 {p_{10}}(x) &=& \frac{-6 \sqrt{3} b^9 \kappa  q^3}{x^4 \left(x^2-1\right) \left(b^6 q^2+1\right)^2 \left(-q^2 b^6 +x^4+x^2\right)} \Big( x^2 \left(2 x^5-3 x^4-2 x+3\right)+
 \\[1ex]
 \nonumber
&+& b^6 q^2 \left(x^7-3 x^6-3 x^3+6 x^2+8 x-7\right)-b^{12} q^4 \left(x^3-3 x^2-4 x+7\right)+
\\[1ex]
\nonumber
&+& 4 x {F_0} \left(b^{12} q^4 \left(3 x^2-4\right)-b^6 q^2 \left(x^6-6 x^2+4\right)-x^2
   \left(x^4-3\right)\right)
  \Big)
 \eea
The existence of terms $1/(1-x^2)$ in $H_1$, $H_3$, $H_4$  may suggest that
there is a problem at the outer horizon. However, examining the near-horizon
behaviour one can verify that each of these functions is regular. To see this
one must make use of the explicit form of $F_2$.

\section{Weyl weights}
\label{appweyl}

Weight $\, \, \, \, \, 4: \quad $ tensors from $T_1^{\mu \nu}$ to $T_{11}^{\mu \nu}$,

Weight $\, \, \, \, \, 3: \quad $  $q$, $\sigma^{\mu \nu}$, $\omega^{\mu \nu}$

Weight $\, \, \, \, \, 2: \quad $  $g^{\mu \nu}$, $l^{\mu}$, ${V_0}^{\mu}$, vectors from $V_1^{\mu} $ to $ V_5^{\mu}$,

Weight $\, \, \, \, \, 1: \quad $  $T$, $\mu$, $u^{\mu}$, $r$,

Weight $\, \, \, \, \, 0: \quad $  $l_{\mu}$, ${V_0}_{\mu}$,  Weyl-invariant scalars, all covariant vectors and tensors

Weight $\, \, \, \, \, -1: \quad $  $b$, $u_\mu$, $\sigma_{\mu \nu}$,
$\omega_{\mu \nu}$.

Weight $\, \, \, \, \, -2: \quad $  $g_{\mu \nu}$.

\section{Inner horizon at second order}
\label{appinnerhor}

The location of the inner horizon is given by \rf{innerhor}. The
coefficient functions appearing there are given by
\bea
\label{innerhorposcoeffs}
 \nonumber
 h_1 &=&  \frac{b^4 r_-^5 {K_1}\left(b r_- \sqrt{b^2 r_-^2+1},b r_-\right)}{-2 b^4 r_-^4+b^2 r_-^2+1}-\frac{b^3 r_-^4 \left(5 b^2
   r_-^2+4\right)}{3 \left(b^2 r_-^2+2\right) \left(-2 b^4 r_-^4+b^2 r_-^2+1\right){}^2},
 \\[1ex]
 \nonumber
 h_2 &=& \frac{1}{60 b^2 r_{-}} \left( \frac{24 \kappa^2 \left(b^2   r_-^2+1 \right)^2}{\left(b^4 r_-^4+b^2 r_-^2+1\right)^2}+\frac{5 \left(-9 b^4  r_-^4+2 b^2 r_-^2+2\right)}{2 b^4 r_-^4-b^2 r_-^2-1}  \right),
   \\[1ex]
   \nonumber
 h_3 &=& \frac{b^2 r_-^3}{12 \left(-2 b^4 r_-^4+b^2 r_-^2+1\right)},  
 \\[1ex]
 \nonumber
 h_4 &=& -\frac{r_-^3 \left(b^3 r_-^2+b\right){}^2}{4 \left(b^2 r_-^2+2\right){}^2 \left(b^4 r_-^4+b^2 r_-^2+1\right) \left(2 b^4
   r_-^4-b^2 r_-^2-1\right){}^3} \Big( -12 b^5 r_-^5+82 b^4 r_-^4-4 b^3 r_-^3+
   \\[1ex]
   \nonumber
  &+& 32 b^2 r_-^2+8+  8 b^{13} r_-^{13}+32 b^{12} r_-^{12}+18 b^{11} r_-^{11}+80 b^{10} r_-^{10}+b^9 r_-^9+130 b^8 r_-^8-11 b^7 r_-^7+
\\[1ex]
\nonumber
 &+& 122 b^6 r_-^6 \Big) + \frac{b^4 r_-^5 {F_0}^{2} \left(b r_- \sqrt{b^2 r_-^2+1},b r_-\right){}}{-4 b^4 r_-^4+2 b^2 r_-^2+2}+\frac{b^4 r_-^5
   {K_4}\left(b r_- \sqrt{b^2 r_-^2+1},b r_-\right)}{-2 b^4 r_-^4+b^2 r_-^2+1}+
\\[1ex]
\nonumber
 &-& \frac{b^3 r_-^4 \left(4 b^6 r_-^6+7 b^4
   r_-^4+5 b^2 r_-^2+2\right) {F_0}\left(b r_- \sqrt{b^2 r_-^2+1},b r_-\right)}{\left(b^2 r_-^2+2\right) \left(-2 b^4
   r_-^4+b^2 r_-^2+1\right){}^2},
    \\[1ex]
    \nonumber
  h_5 &=& \frac{b^4 r_-^5 {K_5}\left(b r_- \sqrt{b^2 r_-^2+1},b r_-\right)}{-2 b^4 r_-^4+b^2 r_-^2+1}+\frac{b^3 r_-^4 \left(4
   b^{10} r_-^{10}+11 b^8 r_-^8+20 b^6 r_-^6+21 b^4 r_-^4+12 b^2 r_-^2+4\right)}{4 \left(-2 b^4 r_-^4+b^2 r_-^2+1\right){}^2
   \left(b^6 r_-^6+3 b^4 r_-^4+3 b^2 r_-^2+2\right)},
   \\[1ex]
\nonumber
 h_6 &=&  \frac{b^4 r_-^5 {K_6}\left(b r_- \sqrt{b^2 r_-^2+1},b r_-\right)}{-2 b^4 r_-^4+b^2 r_-^2+1}+
\\[1ex]
\nonumber
&-& \frac{\sqrt{3} \kappa  r_-
   \left(b^2 r_-^2+1\right){}^{3/2} \left(2 b^4 r_-^4+b^2 r_-^2+1\right) \left(4 b^4 r_-^4+3 b^2 r_-^2+2\right)}{\left(-2 b^4
   r_-^4+b^2 r_-^2+1\right){}^2 \left(b^6 r_-^6+3 b^4 r_-^4+3 b^2 r_-^2+2\right)}
\\[1ex]
\nonumber
 &-& \frac{\sqrt{3} b \kappa  r_-^2 \left(b^2 r_-^2+1\right){}^{3/2} {F_0}\left(b r_- \sqrt{b^2 r_-^2+1},b r_-\right)}{2 b^8
   r_-^8+b^6 r_-^6-2 b^2 r_-^2-1}.
 \eea
In the last formula above, some functions, such as for example ${K_4}\left(b
r_- \sqrt{b^2 r_-^2+1},b r_-\right)$,  are singular. In
fact at the inner horizon, starting from first order, metric and vector potential
are also singular. For instance, the first order function $F_2$ can be
rewritten as follows: 
\bea
\nonumber
F_2 &=&  \int_{b r}^{\infty} \textmd{d}x \frac{x (1+x+x^2)}{(1+x)(-b^6 q^2 + x^2 + x^4)} =
\\
\nonumber
 & & \int_{b r}^{\infty} \textmd{d}x \frac{x (1+x+x^2)}{   (1+x)(x-br_{-})(x+ b r_{-})(1+ b^2 r_{-}^2 + x^2)    }.
\eea
The last expression is singular in the limit  $r \rightarrow r_{-}$, so $F_2$
diverges across the inner horizon. The same holds for many other second-order
functions, like $K_4$ or  $K_5$.

\newpage

\bibliographystyle{utphys}
\bibliography{biblio}{}

\providecommand{\href}[2]{#2}\begingroup\raggedright\begin{thebibliography}{10}

\bibitem{CasalderreySolana:2011us}
J.~Casalderrey-Solana, H.~Liu, D.~Mateos, K.~Rajagopal, and U.~A. Wiedemann,
  ``{Gauge/String Duality, Hot QCD and Heavy Ion Collisions},''
\href{http://arxiv.org/abs/1101.0618}{{\ttfamily arXiv:1101.0618 [hep-th]}}.

\bibitem{Bhattacharyya:2008jc}
S.~Bhattacharyya, V.~E. Hubeny, S.~Minwalla, and M.~Rangamani, ``{Nonlinear
  Fluid Dynamics from Gravity},''
  \href{http://dx.doi.org/10.1088/1126-6708/2008/02/045}{{\em JHEP} {\bfseries
  02} (2008) 045},
\href{http://arxiv.org/abs/0712.2456}{{\ttfamily arXiv:0712.2456 [hep-th]}}.

\bibitem{Janik:2005zt}
R.~A. Janik and R.~B. Peschanski, ``{Asymptotic perfect fluid dynamics as a
  consequence of AdS/CFT},''
  \href{http://dx.doi.org/10.1103/PhysRevD.73.045013}{{\em Phys. Rev.}
  {\bfseries D73} (2006) 045013},
\href{http://arxiv.org/abs/hep-th/0512162}{{\ttfamily arXiv:hep-th/0512162}}.

\bibitem{Banerjee:2008th}
N.~Banerjee, J.~Bhattacharya, S.~Bhattacharyya, S.~Dutta, R.~Loganayagam, {\em
  et al.}, ``{Hydrodynamics from charged black branes},''
  \href{http://dx.doi.org/10.1007/JHEP01(2011)094}{{\em JHEP} {\bfseries 1101}
  (2011) 094}, \href{http://arxiv.org/abs/0809.2596}{{\ttfamily arXiv:0809.2596
  [hep-th]}}.

\bibitem{Erdmenger:2008rm}
J.~Erdmenger, M.~Haack, M.~Kaminski, and A.~Yarom, ``{Fluid dynamics of
  R-charged black holes},''
  \href{http://dx.doi.org/10.1088/1126-6708/2009/01/055}{{\em JHEP} {\bfseries
  0901} (2009) 055}, \href{http://arxiv.org/abs/0809.2488}{{\ttfamily
  arXiv:0809.2488 [hep-th]}}.

\bibitem{Bhattacharyya:2008ji}
S.~Bhattacharyya {\em et al.}, ``{Forced Fluid Dynamics from Gravity},''
  \href{http://dx.doi.org/10.1088/1126-6708/2009/02/018}{{\em JHEP} {\bfseries
  02} (2009) 018},
\href{http://arxiv.org/abs/0806.0006}{{\ttfamily arXiv:0806.0006 [hep-th]}}.

\bibitem{Bhattacharyya:2008mz}
S.~Bhattacharyya, R.~Loganayagam, I.~Mandal, S.~Minwalla, and A.~Sharma,
  ``{Conformal Nonlinear Fluid Dynamics from Gravity in Arbitrary
  Dimensions},'' \href{http://dx.doi.org/10.1088/1126-6708/2008/12/116}{{\em
  JHEP} {\bfseries 12} (2008) 116},
\href{http://arxiv.org/abs/0809.4272}{{\ttfamily arXiv:0809.4272 [hep-th]}}.

\bibitem{Hur:2008tq}
J.~Hur, K.~K. Kim, and S.-J. Sin, ``{Hydrodynamics with conserved current from
  the gravity dual},''
  \href{http://dx.doi.org/10.1088/1126-6708/2009/03/036}{{\em JHEP} {\bfseries
  0903} (2009) 036},
\href{http://arxiv.org/abs/0809.4541}{{\ttfamily arXiv:0809.4541 [hep-th]}}.

\bibitem{Kalaydzhyan:2010iv}
T.~Kalaydzhyan and I.~Kirsch, ``{Holographic dual of a boost-invariant plasma
  with chemical potential},''
  \href{http://dx.doi.org/10.1007/JHEP02(2011)053}{{\em JHEP} {\bfseries 1102}
  (2011) 053},
\href{http://arxiv.org/abs/1012.1966}{{\ttfamily arXiv:1012.1966 [hep-th]}}.

\bibitem{Chamblin:1999tk}
A.~Chamblin, R.~Emparan, C.~V. Johnson, and R.~C. Myers, ``{Charged AdS black
  holes and catastrophic holography},''
  \href{http://dx.doi.org/10.1103/PhysRevD.60.064018}{{\em Phys.Rev.}
  {\bfseries D60} (1999) 064018},
\href{http://arxiv.org/abs/hep-th/9902170}{{\ttfamily arXiv:hep-th/9902170
  [hep-th]}}.

\bibitem{Son:2009tf}
D.~T. Son and P.~Surowka, ``{Hydrodynamics with Triangle Anomalies},''
  \href{http://dx.doi.org/10.1103/PhysRevLett.103.191601}{{\em Phys. Rev.
  Lett.} {\bfseries 103} (2009) 191601},
\href{http://arxiv.org/abs/0906.5044}{{\ttfamily arXiv:0906.5044 [hep-th]}}.

\bibitem{Loganayagam:2008is}
R.~Loganayagam, ``{Entropy Current in Conformal Hydrodynamics},''
  \href{http://dx.doi.org/10.1088/1126-6708/2008/05/087}{{\em JHEP} {\bfseries
  05} (2008) 087},
\href{http://arxiv.org/abs/0801.3701}{{\ttfamily arXiv:0801.3701 [hep-th]}}.

\bibitem{Bhattacharyya:2008xc}
S.~Bhattacharyya {\em et al.}, ``{Local Fluid Dynamical Entropy from
  Gravity},'' {\em JHEP} {\bfseries 06} (2008) 055,
\href{http://arxiv.org/abs/0803.2526}{{\ttfamily arXiv:0803.2526 [hep-th]}}.

\bibitem{Romatschke:2009kr}
P.~Romatschke, ``{Relativistic Viscous Fluid Dynamics and Non-Equilibrium
  Entropy},'' \href{http://dx.doi.org/10.1088/0264-9381/27/2/025006}{{\em
  Class. Quant. Grav.} {\bfseries 27} (2010) 025006},
\href{http://arxiv.org/abs/0906.4787}{{\ttfamily arXiv:0906.4787 [hep-th]}}.

\bibitem{Jensen:2012jh}
K.~Jensen, M.~Kaminski, P.~Kovtun, R.~Meyer, A.~Ritz, {\em et al.}, ``{Towards
  hydrodynamics without an entropy current},''
  \href{http://dx.doi.org/10.1103/PhysRevLett.109.101601}{{\em Phys.Rev.Lett.}
  {\bfseries 109} (2012) 101601},
\href{http://arxiv.org/abs/1203.3556}{{\ttfamily arXiv:1203.3556 [hep-th]}}.

\bibitem{Banerjee:2012iz}
N.~Banerjee, J.~Bhattacharya, S.~Bhattacharyya, S.~Jain, S.~Minwalla, {\em et
  al.}, ``{Constraints on Fluid Dynamics from Equilibrium Partition
  Functions},'' \href{http://dx.doi.org/10.1007/JHEP09(2012)046}{{\em JHEP}
  {\bfseries 1209} (2012) 046},
\href{http://arxiv.org/abs/1203.3544}{{\ttfamily arXiv:1203.3544 [hep-th]}}.

\bibitem{Booth:2009ct}
I.~Booth, M.~P. Heller, and M.~Spalinski, ``{Black brane entropy and
  hydrodynamics: the boost-invariant case},''
  \href{http://dx.doi.org/10.1103/PhysRevD.80.126013}{{\em Phys. Rev.}
  {\bfseries D80} (2009) 126013},
\href{http://arxiv.org/abs/0910.0748}{{\ttfamily arXiv:0910.0748 [hep-th]}}.

\bibitem{Booth:2010kr}
I.~Booth, M.~P. Heller, and M.~Spalinski, ``{Black Brane Entropy and
  Hydrodynamics},'' \href{http://dx.doi.org/10.1103/PhysRevD.83.061901}{{\em
  Phys. Rev.} {\bfseries D83} (2011) 061901},
\href{http://arxiv.org/abs/1010.6301}{{\ttfamily arXiv:1010.6301 [hep-th]}}.

\bibitem{Booth:2011qy}
I.~Booth, M.~P. Heller, G.~Plewa, and M.~Spalinski, ``{On the apparent horizon
  in fluid-gravity duality},''
  \href{http://dx.doi.org/10.1103/PhysRevD.83.106005}{{\em Phys.Rev.}
  {\bfseries D83} (2011) 106005},
\href{http://arxiv.org/abs/1102.2885}{{\ttfamily arXiv:1102.2885 [hep-th]}}.

\bibitem{Plewa:2013rga}
G.~Plewa and M.~Spalinski, ``{Entropy currents from holography in hydrodynamics
  with charge},'' \href{http://dx.doi.org/10.1007/JHEP07(2013)062}{{\em JHEP}
  {\bfseries 1307} (2013) 062},
\href{http://arxiv.org/abs/1304.6705}{{\ttfamily arXiv:1304.6705 [hep-th]}}.

\bibitem{Myers:2009ij}
R.~C. Myers, M.~F. Paulos, and A.~Sinha, ``{Holographic Hydrodynamics with a
  Chemical Potential},''
  \href{http://dx.doi.org/10.1088/1126-6708/2009/06/006}{{\em JHEP} {\bfseries
  06} (2009) 006},
\href{http://arxiv.org/abs/0903.2834}{{\ttfamily arXiv:0903.2834 [hep-th]}}.

\bibitem{Kharzeev:2011ds}
D.~E. Kharzeev and H.-U. Yee, ``{Anomalies and time reversal invariance in
  relativistic hydrodynamics: the second order and higher dimensional
  formulations},'' \href{http://dx.doi.org/10.1103/PhysRevD.84.045025}{{\em
  Phys.Rev.} {\bfseries D84} (2011) 045025},
\href{http://arxiv.org/abs/1105.6360}{{\ttfamily arXiv:1105.6360 [hep-th]}}.

\bibitem{Loganayagam:2011mu}
R.~Loganayagam, ``{Anomaly Induced Transport in Arbitrary Dimensions},''
\href{http://arxiv.org/abs/1106.0277}{{\ttfamily arXiv:1106.0277 [hep-th]}}.

\bibitem{Gauntlett:2006ai}
J.~P. Gauntlett, E.~O~Colgain, and O.~Varela, ``{Properties of some conformal
  field theories with M-theory duals},''
  \href{http://dx.doi.org/10.1088/1126-6708/2007/02/049}{{\em JHEP} {\bfseries
  0702} (2007) 049},
\href{http://arxiv.org/abs/hep-th/0611219}{{\ttfamily arXiv:hep-th/0611219
  [hep-th]}}.

\bibitem{Gauntlett:2007ma}
J.~P. Gauntlett and O.~Varela, ``{Consistent Kaluza-Klein reductions for
  general supersymmetric AdS solutions},''
  \href{http://dx.doi.org/10.1103/PhysRevD.76.126007}{{\em Phys.Rev.}
  {\bfseries D76} (2007) 126007},
\href{http://arxiv.org/abs/0707.2315}{{\ttfamily arXiv:0707.2315 [hep-th]}}.

\bibitem{Witten:1998zw}
E.~Witten, ``{Anti-de Sitter space, thermal phase transition, and confinement
  in gauge theories},'' {\em Adv. Theor. Math. Phys.} {\bfseries 2} (1998)
  505--532,
\href{http://arxiv.org/abs/hep-th/9803131}{{\ttfamily arXiv:hep-th/9803131}}.

\bibitem{DBLP:journals/corr/abs-cs-0608005}
K.~Peeters, ``A field-theory motivated approach to symbolic computer algebra,''
  {\em CoRR} {\bfseries abs/cs/0608005} (2006) .

\bibitem{Peeters:2007wn}
K.~Peeters, ``{Introducing Cadabra: A symbolic computer algebra system for
  field theory problems},''
\href{http://arxiv.org/abs/hep-th/0701238}{{\ttfamily arXiv:hep-th/0701238}}.

\bibitem{Henningson:1998gx}
M.~Henningson and K.~Skenderis, ``{The holographic Weyl anomaly},'' {\em JHEP}
  {\bfseries 07} (1998) 023,
\href{http://arxiv.org/abs/hep-th/9806087}{{\ttfamily arXiv:hep-th/9806087}}.

\bibitem{Balasubramanian:1999re}
V.~Balasubramanian and P.~Kraus, ``{A stress tensor for anti-de Sitter
  gravity},'' \href{http://dx.doi.org/10.1007/s002200050764}{{\em Commun. Math.
  Phys.} {\bfseries 208} (1999) 413--428},
\href{http://arxiv.org/abs/hep-th/9902121}{{\ttfamily arXiv:hep-th/9902121}}.

\bibitem{Gubser:2000ec}
S.~S. Gubser and I.~Mitra, ``{Instability of charged black holes in Anti-de
  Sitter space},''
\href{http://arxiv.org/abs/hep-th/0009126}{{\ttfamily arXiv:hep-th/0009126
  [hep-th]}}.

\bibitem{Bhattacharyya:2007vs}
S.~Bhattacharyya, S.~Lahiri, R.~Loganayagam, and S.~Minwalla, ``{Large rotating
  AdS black holes from fluid mechanics},''
  \href{http://dx.doi.org/10.1088/1126-6708/2008/09/054}{{\em JHEP} {\bfseries
  09} (2008) 054},
\href{http://arxiv.org/abs/0708.1770}{{\ttfamily arXiv:0708.1770 [hep-th]}}.

\end{thebibliography}\endgroup

\end{document}